\documentclass[12pt]{article}
\usepackage{amscd,amssymb,amsmath,latexsym,enumerate}
\usepackage[mathscr]{euscript}
\usepackage{epsfig}
\usepackage{fancybox}
\usepackage{verbatim}
\usepackage{multirow}

\usepackage{tikz}
\usepackage{tikz-cd}
\usepackage{todonotes}
\usepackage{multicol}

\usepackage{xcolor}
\definecolor{MyBlue}{cmyk}{1,0.13,0,0.63}
\definecolor{MyGreen}{cmyk}{0.91,0,0.88,0.52}
\newcommand{\mylinkcolor}{MyBlue}
\newcommand{\mycitecolor}{MyGreen}
\newcommand{\myurlcolor}{black}

\usepackage{hyperref}
\hypersetup{%
  bookmarksnumbered=true,bookmarksopen=false,%
  plainpages=false,
  linktocpage=true,%
  colorlinks=true,breaklinks=true,%
  linkcolor=\mylinkcolor,citecolor=\mycitecolor,urlcolor=\myurlcolor,%
  pdfpagelayout=OneColumn,%
  pageanchor=true,%
}

\usepackage{color}

\title{
Approximate symmetries and conservation laws \\
in topological insulators and associated $\mathbb{Z}$-invariants
}

\author{Nora Doll and Hermann Schulz-Baldes
\\
\\
{\small Department Mathematik, Friedrich-Alexander-Universit\"at Erlangen-N\"urnberg }
\\
{\small Cauerstr. 11, D-91056 Erlangen, Germany}
\\
{\small Email: nora.doll@fau.de, schuba@mi.uni-erlangen.de}
}

\date{ }

\newtheorem{theo}{Theorem}
\newtheorem{defini}{Definition}
\newtheorem{proposi}{Proposition}

\newcommand{\CM}{{\mathbb C}}
\newcommand{\NM}{{\mathbb N}}

\newcommand{\ZM}{{\mathbb Z}}

\newcommand{\Cc}{{\cal C}}

\newcommand{\one}{{\bf 1}}

\newcommand{\Tr}{\mbox{\rm Tr}}

\newcommand{\Ch}{{\rm Ch}} 
 
\newcommand{\Ind}{{\rm Ind}} 
\newcommand{\Ker}{{\rm Ker}} 
 
\newcommand{\Ran}{{\rm Ran}} 
\newcommand{\sgn}{{\rm sgn}} 
\newcommand{\Sig}{{\rm Sig}} 
\newcommand{\diag}{{\rm diag}} 
\newcommand{\od}{{\mbox{\rm\tiny od}}}
\newcommand{\eve}{{\mbox{\rm\tiny ev}}}

\newcommand{\Maj}{{\mbox{\rm\tiny Maj}}}

\begin{document}

\maketitle

\begin{abstract} 
Solid state systems with time reversal symmetry and/or particle-hole symmetry often only have $\mathbb{Z}_2$-valued strong invariants for which no general local formula is known. For physically relevant values of the parameters, there may exist approximate symmetries or almost conserved observables, such as the spin in a quantum spin Hall system with small Rashba coupling. It is shown in a general setting how this allows to define robust integer-valued  strong invariants stemming from the complex theory, such as the spin Chern numbers, which modulo $2$ are equal to the $\mathbb{Z}_2$-invariants. Moreover, these integer invariants can be computed using twisted versions of the spectral localizer. 
\end{abstract}

\section{Introduction}
\label{sec-Intro}

Physical systems can be distinguished by symmetry properties. If one considers merely chiral symmetry (CHS), time-reversal symmetry (TRS) and particle-hole symmetry (PHS), then one obtains the so-called Cartan-Altland-Zirnbauer (CAZ) classes widely used in solid state physics. All insulators within one such CAZ class can further be distinguished by topological invariants which take values either in the integers or just in $\ZM_2=\{0,1\}$. The most stable of such invariants are called strong invariants and their possible values make up Kitaev's periodic table of topological insulators, see Table~1. The periodicity of this table can be explained as a manifestation of Bott periodicity of $K$-theory \cite{Kit}.  As can be seen, there are two diagonals with eight $\ZM_2$-invariants each. While there are index theorems for these $\ZM_2$-invariants \cite{SB,GS,BKR} even in the mobility gap regime, the standard ways to compute them numerically via edge state crossings \cite{ASV,GP} or the spectral localizer \cite{Lor,LS1,LS3} have only been justified in particular situations, even though a rigorous general bulk-boundary correspondence is available \cite{BKR,AMZ}. One of the difficulties is that there is no general cohomological formula for $\ZM_2$-invariants \cite{FMP,Kel,MPT} (other than for the integer-valued entries of the periodic table which are given by Chern numbers and winding numbers, see \cite{PS}). 

\vspace{.2cm}

Furthermore the physical importance of $\ZM_2$-invariants is disputable. Solid state systems are dirty so that conservation laws and symmetry relations may hold only up to error terms (an exception is the particle-hole symmetry resulting merely from the one-particle approximation).  Moreover, there are potentially more interesting and stable invariants taking integer values. This last point is best explained on the particular case of a two-dimensional quantum spin Hall system (case $j=4$ and $d=2$ in Table~1, hence with odd TRS). Historically, non-vanishing $\ZM_2$-invariants were theoretically found in such systems \cite{KM}. However, experiments have clearly shown that conductive properties within the new phase are even stable under perturbations breaking TRS \cite{KRY}. Most likely, the reason for this is the existence of non-vanishing spin Chern numbers introduced by Sheng {\it et al.} \cite{SWSH} and Prodan \cite{Pro,Pro2}. These invariants are connected to an approximately conserved quantity in the physical system, here given by the $z$-component of the spin. Almost conservation means that the commutator of the spin component with the Hamiltonian is small, which reflects that the Rashba spin orbit coupling is small \cite{Pro}. {Furthermore}, there is a rigorous argument showing that { the parity of the spin Chern numbers is equal to the $\ZM_2$-invariant \cite{SB}, that they} lead to conducting edge states \cite{ASV} and that the associated edge currents have stability properties \cite{SB2,MPT}.

\begin{table}
\begin{center}

\begin{tabular}{|c|c|c|c||c||c|c|c|c|c|c|c|c|}
\hline
$j$ & $\!\!$TRS$\!\!$ & $\!\!$PHS$\!\!$ & $\!\!$CHS$\!\!$ & CAZ & $\!d=1\!$ & $\!d=2\!$ & $d\!=3\!$ & $\!d=4\!$ & $\!d=5\!$ & $\!d=6\!$ & $\!d=7\!$ & $\!d=8\!$
\\\hline\hline
$0$ & $+1$&$0$&$0$ & AI & &  & & $2 \, \ZM\;\;\;\;$ & & $\ZM_2$ S & ${\ZM_2}$\, I &  $\ZM\;\;\;\;$
\\
$1$ & $+1$&$+1$&$1$  & BDI & $\ZM\;\;\;\;$ & &  &  & $2 \, \ZM\;\;\;\;$ & & $\ZM_2$ A & $\ZM_2$ A
\\
$2$ & $0$ &$+1$&$0$ & D & ${\ZM_2}$ C & $\ZM\;\;\;\;$ &  & & & $2\,\ZM\;\;\;\;$ &  & $\ZM_2$ Q
\\
$3$ & $-1$&$+1$&$1$  & DIII  & $\ZM_2$ S &  $\ZM_2$ A &  $\ZM$\;\;\;\; &  & & & $2\,\ZM\;\;\;\;$ &
\\
$4$ & $-1$&$0$&$0$ & AII  & &  $\ZM_2$ S & ${\ZM_2}$ \, I & $\ZM\;\;\;\;$ & & & & $2 \, \ZM\;\;\;\;$
\\
$5$ & $-1$&$-1$&$1$  & CII & $2 \, \ZM\;\;\;\;$ &  & $\ZM_2$ \,A & $\ZM_2$ Q & $\ZM\;\;\;\;$ &  & &
\\
$6$ & $0$ &$-1$&$0$ & C&  & $2\,\ZM\;\;\;\;$ &  & $\ZM_2$ Q & ${\ZM_2}$ \, I & $\ZM\;\;\;\;$ & &
\\
$7$ & $+1$&$-1$&$1$  &  CI &  &   & $2 \, \ZM\;\;\;\;$ &  & $\ZM_2\;$ C & $\ZM_2$ C & $\ZM\;\;\;\;$ &
\\
[0.1cm]
\hline
\end{tabular}
\end{center}
\caption{{\it List of the real symmetry classes ordered by {\rm TRS}, {\rm PHS} and {\rm CHS} as well as the {\rm CAZ} label. Then follow the strong invariants in dimension $d=1,\ldots,8$. They constitute Kitaev's periodic table of topological insulators {\rm \cite{Kit}}. The roman letters indicate the approximate law that leads to integer valued strong invariants. Here {\rm I} is an approximate spin inversion symmetry, {\rm S} and {\rm Q} are approximate spin and charge conservation laws, and finally {\rm C} and {\rm A} designate an approximate chiral symmetry and an approximate conservation law, possibly involving spin, a sublattice structure or the particle-hole degree of freedom. All these approximate laws are described in detail in {\rm Section~\ref{sec-RealCAZ}.}  }}
\label{tab-class}
\end{table}

\vspace{.2cm}

The first question addressed in this paper is how to construct integer valued invariants for symmetry classes having {either a vanishing strong invariant or} a $\ZM_2$-invariant according to the periodic table. {Morimoto and Furusaki \cite{MF} and Shiozaki and Sato \cite{SS} showed how this can be achieved by supposing that the system has an extra symmetry or chiral symmetry (which is called an anti-symmetry in \cite{SS}), both of which can either be unitary or anti-unitary. Here it is only supposed that there is an approximately conserved quantity (like the spin component in the quantum spin Hall effect) or an approximate symmetry.} An example for the latter is an approximate chiral symmetry in odd dimensional systems of CAZ class A that was already discussed in \cite{PS}. More precisely, such systems do not have chiral symmetries, but the terms in the Hamiltonian breaking these symmetries are small. {In odd dimension,} this still allows to define integer valued invariants as (higher) winding numbers. Similarly, if an even dimensional system of CAZ class AIII has an approximate conservation law, it can have an associated non-vanishing Chern number. In the two complex cases $(j,d)=(0,0)$ and $(j,d)=(1,1)$, {the strong invariant is $\ZM$-valued, but it may nevertheless vanish and then} an approximate chiral symmetry (C) or approximate conservation (A) may still allow to define non-vanishing integer invariants. {We will also speak simply of an approximate law if either C or A holds. All the above} information is collected in Table~2. The robust simple analysis of these complex cases is carried out in detail in Section~\ref{sec-ComplexCAZ}. {If the conservation law or symmetry is exact and not only approximate (and the systems are periodic), this analysis essentially reduces to that of \cite{MF,SS}.}

\vspace{.2cm}

In principle, this approach readily transposes to all 64 entries of the Kitaev table of real CAZ. However, we decided to focus on the 16 cases with $\ZM_2$-invariants. One reason, {already mentioned above, is that the $\ZM$-invariants associated to the approximate laws may shed light on the physical properties of such topologically non-trivial systems. Another reason} is that it is still a challenge to compute the $\ZM_2$-invariants numerically {and, as will be shown, a considerable simplification results if a suitable approximate law is present.} In fact, the $\ZM_2$-invariant is given by the parity of the integer invariant, just as in the special case of spin Chern numbers \cite{SB} {and similar to the situation in systems with an inversion symmetry \cite{FK} (however, the approximate laws considered here are much more robust than the inversion symmetry).} The analysis of these $\ZM$-invariants corresponding to the $\ZM_2$-entries in the Kitaev table will essentially {appear as} particular applications of the complex {arsenal} described in Section~\ref{sec-ComplexCAZ}. {In the case of exact symmetries in periodic systems, this is again covered by particular cases in \cite{SS}, even though these authors do not establish the connection of the integer invariants with the $\ZM_2$-invariants of the Kitaev table.}

\begin{table}

\begin{center}
\begin{tabular}{|c|c|c|c|c|}
\hline
$j$ &  $\!\!$CHS$\!\!$ & CAZ & $d=0\;$mod$\,2$ & $d=1\;$mod$\,2$ 
\\
\hline\hline
$0$ &$0$& A  & $\ZM\;\;\;\;\;\;$ A $\;\;\;\;\;\;\; $ ${U}$ $\;\;\; \;\;\;\;$ Sec.~\ref{sec-ApproxExchange2} & $\;\;\;\;\;\;\;\;$ C $\;\;\;\;\;\;\;\;\;\;\; $ $\overline{U}$ $\;\;\; \;\;\;$ Sec.~\ref{sec-ApproxChiral} 
 \\
[0.1cm]
\hline
$1$&$ 1$ & AIII & $\;\;\;\;\;\;\;\; $ C,A $\;\;\; $ $\overline{U}_-,U_-\;$ Sec.~\ref{sec-ApproxExchange}  & $\ZM\;\;\;\;\;$ C,A $\;\;\;\;\; $  $\overline{U}_+,U_+\;\;$ Sec.~\ref{sec-ApproxSymChiral}
\\
[0.1cm]
\hline
\end{tabular}
\end{center}
\caption{{\it List of the complex CAZ classes in dimension $d\,\mbox{mod }2$ and the strong invariants without additional approximate symmetry. The {\rm C} and {\rm A} designate an approximate chiral symmetry and an approximate conservation law. {The letters $U,\overline{U},U_\pm,\overline{U}_\pm$ designate the symmetries in the notation of {\rm \cite{SS}} if the approximate laws are exact. For example, $\overline{U}_+$ is an extra chiral symmetry commuting with the symmetry operator from class {\rm AIII}. The table also shows in which section the corresponding case is dealt with.}}}
\label{tab-classComplex}
\end{table}

\vspace{.2cm}

The second contribution of this paper is to show how the invariants {associated to an approximate law} can be computed by a suitable (twisted) modification of the spectral localizer (defined in \cite{LS1,LS3} and Section~\ref{sec-Review} below). Even in the case of quantum spin Hall systems, this is a considerable improvement on numerical procedures to calculate spin Chern numbers \cite{Pro2}. {For some of the $\ZM_2$-invariants there are suggestions to compute them as the sign of the Pfaffian of the spectral localizer \cite{Lor,LS1}, but the connection to the well-known $\ZM_2$-invariants has not been proved and some of the suggestions in \cite{Lor} need modifications.}
The numerical implementation of the twisted spectral localizer is the object of another study.

\vspace{.2cm}

The paper is organized as follows: Section~\ref{sec-Review} reviews earlier results that are needed to state and prove the new contributions of this paper. This includes a description of the index theoretic characterization of the strong invariants for the complex CAZ classes \cite{PS} and of the spectral localizer for integer valued pairings \cite{LS1,LS3}.  Section~\ref{sec-ComplexCAZ} then develops the two main ideas of this paper in the complex CAZ classes. More precisely, it is shown how an approximate chiral symmetry or an approximate conservation law allow to define an integer valued strong invariant and it is shown how to compute this invariant with a twisted spectral localizer. Section~\ref{sec-ReviewReal} then reviews how the strong invariants of the Kitaev table are given in terms of index pairings with real symmetries. This is essentially based on \cite{GS}. The next Section~\ref{sec-RealCAZ} then shows how for each of the $\ZM_2$-entries in Table~1 a specific approximate law allows to construct a strong integer valued invariant which mod $2$ is equal to the given $\ZM_2$-entry. In each case we attempt to argue that the approximate law is physically reasonable and show how to construct a twisted spectral localizer allowing to compute these integer valued strong invariants.

\vspace{.3cm}

\noindent {\bf Acknowledgement:}  This work was partially supported by the DFG. 

\section{Review of prior results on complex CAZ classes}
\label{sec-Review}

\subsection{Hamiltonians and chiral symmetry}
\label{sec-HamChiSym}

Throughout $H=H^*$ will be a Hamiltonian acting on the $d$-dimensional tight-binding Hilbert space $\ell^2(\ZM^d,\CM^L)$. The $L$-dimensional fiber may contain spin, particle-hole, sublattice and other internal degrees of freedom. Moreover, $H$ is supposed to be of finite range or, more generally, have decaying matrix entries in the following sense:

\begin{defini}
\label{def-Local}
A linear operator $A$ on $\ell^2(\ZM^d,\CM^L)$ is said to be local if for all $k\in\NM$ there exists some constant $C_k<\infty$ such that
\begin{equation}
\label{eq-Decay}
\|\langle x|A|y\rangle\|
\;\leq\;\frac{C_k}{1+|x-y|^k}
\;.
\end{equation}
The norm on the l.h.s. is the operator norm on matrices from $\CM^{L\times L}$ and $|x|$ denotes the euclidean length of $x\in\ZM^d$. 
\end{defini}

Locality in one or another form is, of course, a crucial assumption from a physical point of view. Based on the above, it was shown in \cite{ST} that the set of local operators is a $*$-algbra that is invariant under smooth functional calculus. For covariant operators, locality is equivalent to smoothness and these two facts are also well-known (see \cite{PS}).

\vspace{.2cm}

The second assumption on the Hamiltonian is the existence of a spectral gap at the Fermi level. After a shift in energy, the Fermi level can be chosen to be at $0$ so that $H$ is invertible. This implies that the Hamiltonian describes an insulator. For some of the claims below it is sufficient to suppose that $H$ has a mobility gap, but this case will not be dealt with here as it leads to supplementary technical difficulties.  Associated to $H$ is now a Fermi projection 
$$
P\;=\;\chi(H\leq 0)
\;.
$$ 
As $H$ has a gap, $P$ can be written as a smooth function of $H$ and is hence also local. 

\vspace{.2cm}

In the complex cases of the CAZ classification, the Hamiltonian may also have a so-called chiral (or sublattice) symmetry. For the implementation of this symmetry, let us suppose that $H$ acts on $\ell^2(\ZM^d,\CM^{2L})$, that is, the dimension of the fiber is doubled. Further suppose that on $\CM^{2L}$ act the Pauli matrices:
$$
\sigma_1
\;=\;
\begin{pmatrix}
0 & \one \\ \one & 0 
\end{pmatrix}
\;,
\qquad
\sigma_2
\;=\;
\begin{pmatrix}
0 & -\imath\,\one \\ \imath\,\one & 0 
\end{pmatrix}
\;,
\qquad
\sigma_3
\;=\;
\begin{pmatrix}
\one & 0  \\ 0 & -\one  
\end{pmatrix}
\;,
$$
where each entry involves the identity on $\ell^2(\ZM^d,\CM^{L})$. Then $H$ is said to have a chiral symmetry if
\begin{equation}
\label{eq-ChiralSym}
(\sigma_i)^*\, H\,\sigma_i\;=\;-H
\;.
\end{equation}
If $i=3$, then this is equivalent to $H$ being block off-diagonal:
\begin{equation}
\label{eq-ChiralSym2}
(\sigma_3)^*\, H\,\sigma_3\;=\;-H
\qquad
\Longleftrightarrow
\qquad
H
\;=\;
\begin{pmatrix}
0 & A \\ A^* & 0 
\end{pmatrix}
\;.
\end{equation}
Here $A$ is an operator on $\ell^2(\ZM^d,\CM^L)$ which is invertible as $H$ is invertible. If $H$ has a chiral symmetry, then
\begin{equation}
\label{eq-ChiralSym3}
(\sigma_3)^*\, H\,\sigma_3\;=\;-H
\qquad
\Longrightarrow
\qquad
P
\;=\;
\frac{1}{2}
\begin{pmatrix}
\one & -U \\ -U^* & \one 
\end{pmatrix}
\;,
\end{equation}
and $U=A|A|^{-1}$ is called the Fermi unitary \cite{PS}.

\subsection{Topological invariants}
\label{sec-TopInv}

The topological invariant of the Hamiltonian will be given in terms of (Noether) indices of Fredholm operators. There is a by now standard way \cite{PS} to construct them. One considers the $d$-dimensional (unbounded self-adjoint) Dirac operator
\begin{equation} 
\label{eq-DiracOp}
D
\; =\; 
\sum_{i=1}^d X_i\otimes \one_L \otimes \gamma_i  
\,.
\end{equation}
Here $X_1,\ldots,X_d$ are the self-adjoint position operators on $\ell^2(\ZM^d)$ defined by $X_j|x\rangle=x_j|x\rangle$, and $\gamma_1,\ldots,\gamma_d\in\CM^{d'\times d'}$ are anti-commuting self-adjoint matrices of size $d' = 2^{\lfloor \frac{d}{2} \rfloor}$ which square to $\one$, namely they form an irreducible representation of the complex Clifford algebra $\CM_d$ with $d$ generators. If $d$ is even, there exists a symmetry $\gamma\in\CM^{d'\times d'}$ anticommuting with $\gamma_1,\ldots,\gamma_d$, so that $\gamma D\gamma=-D$. Actually, $\gamma=\gamma_{d+1}$ from the representation of $\CM_{d+1}$. The representation can and will be assumed such that $\gamma=\diag(\one,-\one)$ is block diagonal and thus the Dirac operator is off-diagonal in the grading of $\gamma$: 
\begin{equation} 
\label{eq-DiracOpEven}
D
\;=\;
\begin{pmatrix}
0 & D_0 \\ D_0^* & 0
\end{pmatrix}
\;,
\qquad
d\;\mbox{ even}\;.
\end{equation}
Moreover, the identity $\one_L$ in \eqref{eq-DiracOp} acts on the matrix degrees of freedom. Then set
$$
L'\;=\;
\left\{
\begin{array}{cc}
Ld'\;, & d\;\mbox{odd}\;,
\\
L\tfrac{d'}{2}\;, & d\;\mbox{even}\;.
\end{array}
\right.
$$
Then  $D$ acts on $\ell^2(\ZM^d,\CM^{L'})$ for $d$ odd, while for even $d$, $D$ acts $\ell^2(\ZM^d,\CM^{2L'})$ and $D_0$ on $\ell^2(\ZM^d,\CM^{L'})$. As the commutator of $D$ with any local operator is bounded, the Dirac operator specifies an even or odd Fredholm module for the algebra of local operators. Hence one has standard index pairings that are described next. From the data of the Dirac operator, one can define the Hardy projection
$$
E\;=\;\chi(D\geq 0)
\;,
$$
and for $d$ even the (unitary) Dirac phase $F$ by
\begin{equation} 
\label{eq-DiracPhase}
2\,E-\one
\;=\;
\begin{pmatrix}
0 & F \\ F^* & 0
\end{pmatrix}
\;,
\qquad
d\;\mbox{ even}\;.
\end{equation}
From $E$ and $F$ and the data of the Hamiltonian given by the Fermi projection or Fermi unitary (if $H$ has a chiral symmetry) one now defines the index pairings as the operator
\begin{equation}
\label{eq-ComplexPairingDef}
T
\;=\;
\left\{
\begin{array}{cc}
PFP\,+\one-P\;, &d\; \mbox{\rm even}\;,
\\
E\,UE\,+\,\one-E\;, &  \;\;\;\;\;\;\;\;\;\; \;\;\;\;\;\;\;\;\;\; d\; \mbox{\rm odd and }H\;\mbox{\rm chiral}\;.
\end{array}
\right.
\end{equation}
Then $T$ is a Fredholm operator, namely it has finite dimensional kernel and cokernel. The associated index 
$$
\Ind(T)
\;=\;
\dim\big(\Ker(T)\big)\,-\,\dim\big(\Ker(T^*)\big)
\;,
$$ 
is called the (strong) topological invariant. It takes values in $\ZM$ and provides the entries in Table~2.  Admittedly, this may seem like an awkward way to introduce topological invariants. However, if the Hamiltonian is part of a covariant family of Hamiltonians \cite{PS}, then the index is almost surely constant and equal to either Chern numbers (for even $d$) or higher winding numbers (for odd $d$, also called odd Chern number): 
{
\begin{equation}
\label{eq-ComplexPairingDef2}
\Ind(T)
\;=\;
\left\{
\begin{array}{cc}
\Ch_d(P)\;, &d\; \mbox{\rm even}\;,
\\
\Ch_d(U)\;, &  \;\;\;\;\;\;\;\;\;\; \;\;\;\;\;\;\;\;\;\; d\; \mbox{\rm odd and }H\;\mbox{\rm chiral}\;.
\end{array}
\right.
\end{equation}
Here the non-commutative formulas for the Chern numbers on the r.h.s. reduce to standard definitions of the strong invariants \cite{PS}.} The equalities \eqref{eq-ComplexPairingDef2} are the statement of index theorems { that even hold if the Fermi level lies merely in a mobility gap}  \cite{PS}. One of the advantages of {using the index directly as the strong invariant} is that it does not require covariant Hamiltonians and can thus be associated to a single Hamiltonian. The topological content of the index can then be read off the spectral flow upon the insertion of a monopole \cite{CSB}. For the present paper, the index approach is technically advantageous.

\subsection{The spectral localizer}
\label{sec-SpecLoc}

The spectral localizer is another Dirac-like operator combining the Hamiltonian from Section~\ref{sec-HamChiSym} with the Dirac operator from Section~\ref{sec-TopInv}, just as the index pairing \eqref{eq-ComplexPairingDef}. The main interest in this object is that its finate dimensional approximations have a spectral asymmetry that is equal to the invariant given by the index pairing, see Theorem~\ref{theo-SpecLoc} below. Let us describe the construction and result in some detail, as one of the aims of this paper is to modify the spectral localizer. As for the index pairings, the odd and even cases have to be treated separately. For $d$ odd, the Dirac operator $D$ acts on $\ell^2(\ZM^d,\CM^{L'})$ where $L'=Ld'$. On the other hand, a chiral Hamiltonian $H$ acting on $\ell^2(\ZM^d,\CM^{2L})$ is  in the form \eqref{eq-ChiralSym2} with matrix entries $A$ acting on $\ell^2(\ZM^d,\CM^{L})$. They are identified with $H\otimes\one$ and $A\otimes \one$ acting on $\ell^2(\ZM^d,\CM^{2L'})$ and $\ell^2(\ZM^d,\CM^{L'})$ respectively. For a tuning parameter $\kappa>0$, the spectral localizer is then defined to be the operator 
\begin{equation}
\label{eq-OddSpecLoc}
L^\od_\kappa 
\;=\;
 \begin{pmatrix} \kappa\,D & A \\ A^* & -\kappa\,D\end{pmatrix}
 \;=\;
\kappa\,D\otimes\sigma_3
\,+\, 
H
\;,
\end{equation}
acting on $\ell^2(\ZM^d,\CM^{2L'})$. For $d$ even, with $D_0$ acting on $\ell^2(\ZM^d,\CM^{L'})$ and given by \eqref{eq-DiracOpEven} and $H\cong H\otimes\one$ also acting on $\ell^2(\ZM^d,\CM^{L'})$, the spectral localizer is defined again as an operator on $\ell^2(\ZM^d,\CM^{2L'})$ by
\begin{equation}
\label{eq-EvenSpecLoc}
L^\eve_\kappa 
\;=\; 
\begin{pmatrix} H & \kappa \,D_0 \\ \kappa\, D_0^* & -H\end{pmatrix} 
\;=\; \kappa \,D\,+\, H \otimes \gamma
\;.
\end{equation}
Let us note that both $L^\od_\kappa $ and $L^\eve_\kappa $ are self-adjoint (domain issues are not discussed here).
Now the finite volume restriction of the spectral localizer are constructed using the partial isometry $\pi_\rho$ from $\ell^2(\ZM^d,\CM^{L'})$ onto $\Ran(\chi(|D|\leq \rho))$ and $\Ran(\chi(|D_0|\leq \rho))$, respectively for odd and even $d$. Here $\rho>0$ is a diameter (or spectral radius of $D$). As $D$ has compact resolvent, the range of $\pi_\rho$ is finite dimensional. Note that, for even $d$, the operator $D_0$ is normal so that $|D_0|=|D_0^*|$. It is now natural to consider also $\pi_\rho\oplus\pi_\rho$ which will also simply be denoted by $\pi_\rho$. Now for any operator $B$ on $\ell^2(\ZM^d,\CM^{L'})$ or $\ell^2(\ZM^d,\CM^{2L'})$, the finite volume restriction is defined by $B_\rho=\pi_\rho B\pi_\rho^*$. The finite volume spectral localizer is then defined by
$$
L^\od_{\kappa,\rho}\;=\;(L^\od_\kappa)_\rho
\;=\;
 \begin{pmatrix} \kappa\,D_\rho & A_\rho \\ A^*_\rho & -\kappa\,D_\rho\end{pmatrix}
\;,
\qquad
L^\eve_{\kappa, \rho} 
\;=\; 
(L^\eve_{\kappa})_{ \rho} 
\;=\; 
\begin{pmatrix} H_\rho & \kappa \,D_{0,\rho} \\ \kappa\, D_{0,\rho}^* & -H_\rho \end{pmatrix}
\;.
$$
These are both finite dimensional self-adjoint matrices. In the following, the upper index on $L^\od_{\kappa, \rho}$ and $ L^\eve_{\kappa, \rho}$ will be dropped.

\begin{theo}[\cite{LS2,LS3}]
\label{theo-SpecLoc}
Let $g$ be the invertibility gap of $H$, namely $g=\|H^{-1}\|^{-1}$. For odd $d$ suppose the $H$ is of the form \eqref{eq-ChiralSym2}. For odd and even $d$ respectively, suppose that the tuning parameter $\kappa$ and radius $\rho$ are admissible in the sense that the bounds
\begin{equation}
\label{eq:kappa}
\kappa 
\;\leq \;
\frac{g^3}{12 \left\|H\right\| \left\|\left[D,A\right]\right\|}
\;,
\qquad
\kappa 
\;\leq \;
\frac{g^3}{12 \left\|H\right\| \left\|\left[D_0,H\right]\right\|}
\;,
\end{equation}
and
\begin{equation}
\label{eq:rho}
\rho\;>\;\frac{2g}{\kappa}
\end{equation}
hold. Then
\begin{equation}
\label{eq-LBound}
(L_{\kappa, \rho})^2 
\;\geq \;
\frac{g^2}{4} \,\one _\rho
\;,
\end{equation}
namely $L_{\kappa, \rho}$ is invertible so that its signature is well-defined. Furthermore, for the index pairings $T$ given by \eqref{eq-ComplexPairingDef}, one has 
$$
\Ind (T) \;=\; \frac{1}{2} \;\Sig  \left( L_{\kappa, \rho} \right)
\;.
$$
\end{theo}

The main interest of the localizer is that it allows to access the topological invariants without heavy numerical computations. Indeed, the finite-dimensional matrix $L_{\kappa,\rho}$ is merely built from the matrix entries of the Hamiltonian and the position operator. No spectral calculus of the Hamiltonian is needed (namely, one does not need to compute the Fermi projection), not even of the spectral localizer itself because one can compute the signature directly with the block Chualesky decomposition. As stressed in \cite{LS2,LS3} as well as \cite{LS1}, the conditions \eqref{eq:kappa} and \eqref{eq:rho} are not optimal, but are likely not far from optimal. On the other hand, even if $H$ has merely a mobility gap, numerics have shown that the (fluctuation) signature of the spectral localizer is still linked to the strong invariant  (see \cite{LSS}).

\section{Complex CAZ classes with approximate symmetries}
\label{sec-ComplexCAZ}

In this section, approximate symmetries and the associated invariants are discussed for the complex CAZ classes (hence no real structure is involved). Let us first give a short overview by discussing the entries of Table~\ref{tab-classComplex}. For $d$ even, there is a $\ZM$-valued strong invariant for systems without symmetry, namely CAZ class A. It is known to be equal to the Chern numbers \cite{PS}. For odd $d$, there is a  $\ZM$-valued strong invariant for systems with a chiral symmetry. This invariant is known to be equal to the (higher) winding numbers \cite{PS}. Now for odd $d$ and CAZ class A, the {standard periodic} table has no entry, indicating that there is no non-trivial strong invariant. If there is, however, an approximate chiral symmetry then it was already noted in \cite{PS} and will be further discussed in Section~\ref{sec-ApproxChiral} that the strong invariant from odd-dimensional chiral systems is still well-defined.  Similarly, for even dimensional chiral systems an approximate law still allows to define a strong invariant, see Section~\ref{sec-ApproxExchange}. Furthermore, in the two cases with possible strong invariant, namely $(j,d)=(1,1)$ and $(j,d)=(0,0)$,  the strong invariant may vanish. However, if these systems then have a further approximate law (either chiral symmetry or conservation law, see Table~2), then there may nevertheless be again non-vanishing strong invariants, see Sections~\ref{sec-ApproxExchange2} and \ref{sec-ApproxSymChiral} {respectively}. {This analysis hence extends the works \cite{MF} and \cite{SS} on the complex classes to the case of approximate laws.} Let us also note that, in principle, one can iterate the procedure. For example, if  for $(j,d)=(0,1)$ the approximate chiral symmetry only leads to a vanishing invariant, one may look for a further approximate symmetry. This is not investigated here though.

\vspace{.2cm}

After having defined the strong invariants associated to approximate laws, the second aim of this section is to show how a twisted modification of the spectral localizer allows to determine the invariant in an efficient manner,  which is also susceptible to numerical evaluation.

\subsection{Approximate chiral symmetry {without other symmetry}}
\label{sec-ApproxChiral}

Let us begin with the case of an odd dimensional system described by a Hamiltonian $H$ on $\ell^2(\ZM^d,\CM^{2L})$ without chiral symmetry, but for which
$$
\eta
\;=\;
\|\sigma_3 H\sigma_3\,+\,H\|
$$
is small. This will be called an approximate chiral symmetry. Strictly speaking, $H$ is not in CAZ class AIII and hence according to the Kitaev periodic table there would not be any strong invariant associated with $H$, but as in \cite{PS} one can readily argue that the $\ZM$-valued invariant can still exist. Indeed, if $H$ with approximate chiral symmetry is written in the grading of the Pauli matrices as
\begin{equation}
\label{eq-HEntries}
H
\;=\;
\begin{pmatrix}
H_+ & A \\ A^* & H_-
\end{pmatrix}
\;,
\end{equation}
then $\eta=2\,\max\{\|H_+\|,\|H_-\|\}$. If $g=\|H^{-1}\|^{-1}$, then $\eta<2g$ assures that $A$ is invertible.  Hence from the operator $A$ one can construct the odd index pairing from the CAZ class AIII and in odd dimension it can take arbitrary integer values. Why is this of any use? First of all, the index allows to distinguish different topological ground states. Second of all, the bulk-boundary correspondence is still valid in an approximate manner. This is best illustrated in dimension $d=1$ for which the topological systems in class AIII are then given by stacking Su-Schrieffer-Heeger models. The integer index pairing is then equal to the number of zero-energy chiral edge modes for a half-line system ({\it e.g.} \cite{PS}). If now $H_+$ and $H_-$ are added, these zero modes will be shifted away from $0$, but for small enough $\eta$ they can still be found in the gap of the bulk spectrum. Hence in a weak perturbative sense, the bulk-boundary correspondence persists. In the following, the bulk-boundary correspondence will not be discussed for the other approximate symmetries, but we do expect similar weak forms for all of them.

\vspace{.2cm}

The second focus of this paper is rather on showing how a modification of the odd spectral localizer allows to determine the index in an efficient manner,  also in numerical computations. This modification of \eqref{eq-OddSpecLoc} is given by
\begin{equation}
\label{SpecLocOddMod}
L_\kappa 
\;=\;
\begin{pmatrix}\kappa\,\sigma_3\,D & H \\ H & \kappa\,\sigma_3\,D\end{pmatrix}
\;.
\end{equation}
Here $\sigma_3$ is a "twist" of the Dirac operator and such twist will appear in other spectral localizers below. The proof of the following result is then deferred to the Appendix~\ref{app-SpecLocProof}.

\begin{proposi}
\label{prop-SpecLocApproxChiral}
Let $d$ be odd and $g=\|H^{-1}\|^{-1}$. Suppose that $H$ is of the form {\eqref{eq-HEntries}}. Let the tuning parameter satisfy 
\begin{equation}
\label{eq:kappa2}
\kappa 
\;\leq \;
\frac{2g^3}{81 \left\|H\right\| \left\|\left[D,A\right]\right\|}
\;,
\end{equation}
and the radius $\rho$ and the perturbation size $\eta$ satisfy
\begin{equation}
\label{eq:rho2}
\rho\;>\;\frac{8g}{3\kappa}
\;,
\qquad
\eta\;<\;\frac{2g}{3}
\;.
\end{equation}
Then, { with the spectral localizer defined by \eqref{SpecLocOddMod},}
\begin{equation}
\label{eq-IndSigMod}
\Ind (EA|A|^{-1}E+\one-E) \;=\; \frac{1}{4} \;\Sig  \left( L_{\kappa, \rho} \right)
\;.
\end{equation}
\end{proposi}

Let us stress that, as in the cases described in Section~\ref{sec-SpecLoc}, the conditions \eqref{eq:kappa2} and \eqref{eq:rho2} are sufficient for \eqref{eq-IndSigMod} to hold, but the r.h.s. may well be of interest in more general situations, in particular, even in a mobility gap regime. A possible modification of \eqref{SpecLocOddMod} is to work with a spectral localizer
$$
\widetilde{L}_{\kappa,\rho}
\;=\;
\begin{pmatrix} H_++\kappa\,D & A \\ A^* & H_- -\kappa\,D\end{pmatrix}
\;.
$$
Still one has for admissible $(\kappa,\rho)$ that 
$$
\Ind (EA|A|^{-1}E+\one-E) \;=\; \frac{1}{2} \;\Sig  \left( \widetilde{L}_{\kappa, \rho} \right)\;.
$$
The matrix dimension of $\widetilde{L}_{\kappa,\rho}$ is by a factor $2$ smaller than that of ${L}_{\kappa,\rho}$, which is advantageous for numerics. On the other hand, the twisted version \eqref{SpecLocOddMod} is more intuitive.

\subsection{Approximate conservation law {without other symmetry}}
\label{sec-ApproxExchange2}

{This section is tailored for even-dimensional systems without symmetry. They may have a non-vanishing strong invariant, but most interesting is the case where this invariant vanishes and the system nevertheless may have non-trivial topology. This can result from} a conservation law of a Hamiltonian $H$ which is a commutation relation of the form
\begin{equation}
\label{eq-ExchangeSym}
[H,\sigma_i]\;=\;0
\;,
\end{equation}
where $\sigma_i$ is one of the Pauli matrices $\sigma_1,\sigma_2,\sigma_3$. This requires that $H$ acts on $\ell^2(\ZM^d,\CM^{2L})$ with an even dimensional fiber on which again the Pauli matrices act. An approximate conservation law then requires
$$
\eta
\;=\;
\|(\sigma_i)^*\,H\,\sigma_i\,-\,H\|
\;=\;
\|[H,\sigma_i]\|
$$
to be small.  Note the difference of sign in \eqref{eq-ExchangeSym} w.r.t. the chiral symmetry \eqref{eq-ChiralSym}. In a spinful system, $\sigma_i$ could be a component of a spin and as such it appears later on ({\it e.g.} Section~\ref{sec-SQHE} dealing with the quantum spin Hall effect). Furthermore, for a BdG operator it can represent the charge operator, see Section~\ref{sec-HamSym}. Finally let us add that one could also require the approximate conservation of a self-adjoint operator other than a Pauli matrix $\sigma_i$; if the observable is gapped, one can then consider the symmetry built from the projection below the gap. This slight generalization is not spelled out in detail here. If now \eqref{eq-ExchangeSym} holds with $i=3$, the conservation law is equivalent to $H$ being block diagonal, namely if the block entries of $H$ are as in \eqref{eq-HEntries}, then
\begin{equation}
\label{eq-ExchangeSym2}
(\sigma_3)^*\, H\,\sigma_3\;=\;H
\qquad
\Longleftrightarrow
\qquad
H
\;=\;
\begin{pmatrix}
H_+ & 0 \\ 0 & H_- 
\end{pmatrix}
\;,
\end{equation}
with $H_\pm$ being self-adjoint operators on $\ell^2(\ZM^d,\CM^{L})$. In view of \eqref{eq-ExchangeSym2}, an approximate conservation law with $\sigma_3$ allows to split the system into a sum of two subsystems, up to errors. More precisely,  with matrix entries as in \eqref{eq-HEntries}, one has $\eta=2\|A\|$  and then
\begin{equation}
\label{eq-DiagHamInv}
\begin{pmatrix}
H_+ & 0 \\ 0 & H_- 
\end{pmatrix}
\;=\;
H\left[
\one\,-\,
H^{-1}\,
\begin{pmatrix}
0 & A \\ A^* & 0 
\end{pmatrix}
\right]
\end{equation}
and a bound $\eta<2g$ with $g=\|H^{-1}\|^{-1}$ implies by a Neumann series that $\diag(H_+,H_-)$ is invertible. Then both $H_\pm$ have a spectral projection $P_\pm=\chi(H_\pm<0)$ which is local. Thus in even dimensions there are two invariants $\Ind(P_\pm FP_\pm+\one-P_\pm)$. Considering the homotopy
$$
t\in[0,1]\;\mapsto\;
H(t)
\;=\;
\begin{pmatrix}
H_+ & t\,A \\ t\,A^* & H_-
\end{pmatrix}
$$
inside the gapped local operators, one has a path  $t\in[0,1]\mapsto P(t)=\chi(H(t)<0)$ of local projections, thus of constant index so that
\begin{equation}
\label{eq-SumIndex}
\Ind(P_+ FP_++\one -P_+)
\;+\;
\Ind(P_- FP_-+\one -P_-)
\;=\;
\Ind(P FP+\one -P)
\;.
\end{equation}
Hence even if the standard strong invariant $\Ind(P FP+\one -P)$ vanishes, one can have non-trivial strong invariants $\Ind(P_\pm FP_\pm+\one-P_\pm)$ adding up to $0$. The spectral localizer suitable for the computation of these invariants is 
\begin{equation}
\label{eq-SpecLocConsLaw}
L_\kappa
\;=\;
\begin{pmatrix}
H & \kappa \, D \, \sigma_1
\\
\kappa\,\sigma_1 \,D & -H
\end{pmatrix}
\;.
\end{equation}
Again the proof of the following result is given in Appendix~\ref{app-SpecLocProof}.

\begin{proposi}
\label{prop-SpecLocApproxConsLaw}
Let $d$ be even and $g=\|H^{-1}\|^{-1}$.  Suppose that the tuning parameter satisfies 
\begin{equation}
\label{eq:kappa3}
\kappa 
\;\leq \;
\frac{2g^3}{81 \left\|H\right\| \max\{\left\|\left[D,H_+\oplus H_+\right]\right\|,\left\|\left[D,H_-\oplus H_-\right]\right\|\}}
\;,
\end{equation}
and that $\rho$ and $\eta$ satisfy \eqref{eq:rho2}. Then 
$$
\Ind(P_+ FP_++\one -P_+)
\;-\;
\Ind(P_- FP_-+\one -P_-)
\;=\;
\frac{1}{2}\;
\Sig(L_{\kappa,\rho})
\;.
$$
\end{proposi}

Following Prodan \cite{Pro}, another way to split $P$ is to consider $P\sigma_3P$ as a self-adjoint operator on the range of $P$. If $[H,\sigma_3]=0$, the spectrum of $P\sigma_3 P$ is equal to $\{-1,1\}$ and has, in particular, a spectral gap at $0$. Now due to the existence of the gap of $H$, $P$ can be written as a contour integral along a path $\Gamma$ which has minimal distance $g$ from the spectrum of $H$. Hence
$$
[\sigma_3,P]
\;=\;
\oint_{\Gamma} \frac{dz}{2\pi \imath} 
\;[\sigma_3,(z-H)^{-1}]
\;=\;
\oint_{\Gamma} \frac{dz}{2\pi \imath} 
\;(z-H)^{-1}[H,\sigma_3](z-H)^{-1}
\;,
$$
so that $\|[\sigma_3,P]\|\leq Cg^{-2} \eta$ for a constant $C$ that is essentially the length of the path (roughly the size of the spectrum). If now $Cg^{-2} \eta<1$, one concludes that $0$ is not in the spectrum of $P\sigma_i P$ because $(P\sigma_3 P)^2=P\big(\one -(\imath[\sigma_3,P])^2\big)P$. Then one can set $Q_{\pm}=\chi\big(\pm P\sigma_3 P>0\big)$ which is then a local projection and has,  in even dimensions, a well-defined index $\Ind(Q_\pm F Q_\pm+\one-Q_\pm)$. Now the above $P_\pm$ are not equal  to $Q_\pm$, however, they are homotopic to each other within the set of local projections. One has $P=Q_++Q_-$, but in general $P\not=P_++P_-$. Let us stress though that the condition $\eta<g^{2}C^{-1}$ needed to define $Q_\pm$ is considerably more stringent than $\eta<2g$ needed for the definition of $P_\pm$.

\subsection{{Anti-commuting approximate laws for chiral systems}}
\label{sec-ApproxExchange}

This section deals with even dimensional systems with a chiral symmetry {which also have either (i) an approximate conservation law or (ii) an approximate chiral symmetry. Moreover, the two associated symmetry operators are supposed to be anticommuting. On first sight, these may appear to be awkward situations to consider, but they turn out to be relevant for several BdG systems, {\it e.g.} the CAZ class $(j,d)=(3,2)$ treated in Section~\ref{sec-3,2}. If the chiral symmetry takes the form $\sigma_3 H\sigma_3=-H$ as in \eqref{eq-ChiralSym2}, one would hence suppose that the Hamiltonian satisfies 
$$
\mbox{(i) }\; 
\eta\;=\;\|\sigma_1H\sigma_1\,+\,H\|
\;\;\mbox{\rm small}
\;,
\qquad
\mbox{(ii) }\; 
\eta\;=\;\|[H,\sigma_1]\|
\;\;\mbox{\rm small}\,.
$$ 
These two cases are related because one can pass from on to the other. For example, if (i) is given, then considering the symmetry operator $\sigma_3\sigma_1=\imath\sigma_2$ anti-commuting with $\sigma_3$ one gets $\eta=\|[H,\sigma_2]\|$ and is hence in case (ii). Let us focus here on case (ii) as it is the only one that is relevant in Section~\ref{sec-RealCAZ}. It is more convenient to conjugate the equations by the eigenbasis of $\sigma_1$, namely to consider
\begin{equation}
\label{eq-ChiralApproxEx}
\mbox{(ii) }\; 
\sigma_1 H\sigma_1\,=\,-\,H
\;,
\qquad
\eta\;=\;\|[H,\sigma_3]\|
\;\;\mbox{\rm small}\,.
\end{equation}
Writing $H$ then in the block form \eqref{eq-HEntries}, this becomes
\begin{equation}
\label{eq-ChiralApproxEx2}
\mbox{(ii) }\;
H\;=\;
\begin{pmatrix}
H_+ & A \\ -A & -H_+
\end{pmatrix}
\;,
\qquad
\eta\;=\;2\,\|A\|
\;\;\mbox{\rm small}\,.
\end{equation}
}
Arguing as in \eqref{eq-DiagHamInv} one concludes that $H_+$ is invertible if $\eta<2\|H^{-1}\|^{-1}$. Hence there is a local projection $P_+=\chi(H_+\leq 0)$ which in even dimensions has an invariant $\Ind(P_+FP_++\one -P_+)$. The twisted spectral localizer suitable for the computation of this index is
\begin{equation}
\label{eq-ChiralApproxExSpecLoc}
L_\kappa
\;=\;
\begin{pmatrix}
H & \kappa \, D \,\sigma_1
\\
\kappa\,\sigma_1 \, D & -H
\end{pmatrix}
\end{equation}
Then by Proposition~\ref{prop-SpecLocApproxConsLaw}, 
$$
\Ind(P_+ FP_++\one -P_+)
\;=\;
\frac{1}{4}\;
\Sig(L_{\kappa,\rho})
\;,
$$
provided that $\kappa \leq  2g^3\big( 81 \|H\| \|[D,H_+\oplus H_+]\|\big)^{-1}$ and \eqref{eq:rho2} holds.

\subsection{{Commuting approximate laws for chiral systems}}
\label{sec-ApproxSymChiral}

Section~\ref{sec-ApproxExchange2} considered an even-dimensional system without symmetry, namely $(j,d)=(0,0)$, for which the standard $\ZM$-invariant is well-defined, but vanishes. Given an approximate conservation law, it is nevertheless possible to write this as a sum of two potentially non-vanishing invariants, see \eqref{eq-SumIndex}. A similar situation may also appear for odd-dimensional chiral systems, that is, $(j,d)=(1,1)$. The supplementary approximate law, either a second chiral symmetry or a conservation law, will be supposed to commute with the chiral symmetry $\sigma_3H\sigma_3=-H$. Hence it will be implemented by a second set of Pauli matrices $\nu_1,\nu_2,\nu_3$ commuting with the $\sigma_i$ so that the Hamiltonian is supposed to act on $\ell^2(\ZM^d,\CM^{4L})$ to accommodate both the $\nu_i$ and $\sigma_i$. Now let us suppose smallness of either
$$
\mbox{(i) }\; 
\eta
\;=\;
\|\nu_3 \,H\,\nu_3\;+\;H\|
\qquad
\mbox{ or } 
\qquad
\mbox{(ii) } \;
\eta
\;=\;
\|\nu_3 \,H\,\nu_3\;-\;H\|
\;.
$$
{Again these two cases are related. If (i) is given, then considering the symmetry operator $\sigma_3\nu_3$ commuting with $\sigma_3$, one gets $\eta=\|[H,\sigma_3\nu_3]\|$ and is hence in case (ii), and vice versa. Nevertheless, let us treat both cases in some detail because they both appear naturally in Section~\ref{sec-RealCAZ}.} As $H$ is chiral, it is of the form \eqref{eq-ChiralSym2}. In the grading of the $\nu_i$, the operator $A$ on $\ell^2(\ZM^d,\CM^{2L})$ can now be decomposed as
\begin{equation}
\label{eq-AEntries}
A
\;=\;
\begin{pmatrix}
A_+ & B \\ C & A_-
\end{pmatrix}
\;.
\end{equation}
Then the two above cases become
$$
\mbox{(i) }\; 
\eta
\;=\;
2\;\max\{\|A_+\|,\|A_-\|\}
\qquad
\mbox{ or } 
\qquad
\mbox{(ii) } \;
\eta
\;=\;
2\;\max\{\|B\|,\|C\|\}
\;,
$$
and the odd index pairings become, provided that $\eta<2g$,
\begin{align*}
&
\mbox{(i) }\; \;
\Ind(EB|B|^{-1}E+\one-E)\;=\;-\,\Ind(EC|C|^{-1}E+\one-E)\;,
\\
&
\mbox{(ii) } \;
\Ind(EA_+|A_+|^{-1}E+\one-E)\;=\;-\,\Ind(EA_-|A_-|^{-1}E+\one-E)
\;.
\end{align*}
{In both cases, one has  $\Ind(EA|A|^{-1}E+\one-E)=0$, but similarly as in  \eqref{eq-SumIndex} this results from a pair of indices summing up to $0$.} Provided that conditions similar to those in Propositions~\ref{prop-SpecLocApproxChiral} and \ref{prop-SpecLocApproxConsLaw} hold, these invariants can be computed with the following spectral localizers:
\begin{align}
&
\mbox{(i) }\; \;
\Ind(EB|B|^{-1}E+\one-E)\;=\;\frac{1}{4}\,\Sig(L_{\kappa,\rho})
\;,
\qquad
\;\;\;\;
L_\kappa
\;=\;
\begin{pmatrix}
\kappa \, \nu_3\,D & A
\\
A^* & \kappa\,\nu_3\,D
\end{pmatrix}
\;,
\label{eq-IndexPairLocCase(i)}
\\
&
\mbox{(ii) } \;
\Ind(EA_+|A_+|^{-1}E+\one-E)\;=\;\frac{1}{4}\,\Sig(L_{\kappa,\rho}) 
\;,
\qquad
L_\kappa
\;=\;
\begin{pmatrix}
\kappa \, \nu_3\,D & A
\\
A^* & -\,\kappa\,\nu_3\,D
\end{pmatrix}
\;.
\label{eq-IndexPairLocCase(ii)}
\end{align}
%

\section{Review of prior results on real CAZ classes}
\label{sec-ReviewReal}

In the CAZ classification the Hamiltonian can have further symmetries, namely a particle-hole symmetry leading to Bogoliubov-de Gennes operators or a time-reversal symmetry. Both are real symmetries invoking a complex conjugation. All this is recalled in Section~\ref{sec-HamSym}. Similarly, the Dirac operator has real symmetries depending on the dimension. This is inherited from the real Clifford representation and reviewed in Section~\ref{sec-DiracSym}. The combination of these symmetries have implications on the index pairings. Particular focus will be on $\ZM_2$-valued index pairings, see Section~\ref{sec-Z2Index}. Finally Section~\ref{sec-SpecLocSym} discusses real symmetries of the spectral localizer.

\subsection{Symmetries of the Hamiltonian}
\label{sec-HamSym}

The Hamiltonian $H=H^*$ is supposed to be as in Section~\ref{sec-HamChiSym}, namely it acts on a $d$-dimensional tight-binding Hilbert space with finite dimensional fiber and satisfies the locality bound \eqref{eq-Decay}. The Hilbert space is $\ell^2(\ZM^d,\CM^{lL})$ with $l=2,4,8$ so that the action of several sets of Pauli matrices $s_i$, $\tau_i$ and $\nu_i$, commuting with each other, can be implemented. We will not keep track of the size index $l$ explicitly in the following, hoping that the reader has little difficulties to determine it. Furthermore, the Hilbert space is equipped with an anti-unitary involution $\Cc$ called complex conjugation. It is simply defined fiberwise. For any operator $A$, we will also denote its complex conjugate by $\overline{A}=\Cc A\Cc$. It is a linear operator. Also let us denote the transpose of $A$ by $A^T=\Cc A^*\Cc$. The time-reversal symmetry (TRS) will be implemented by Pauli matrices $s_i$ acting on spin degrees of freedom. A Hamiltonian is said to have TRS if
\begin{equation}
\label{eq-TRS}
(s_i)^*\,\overline{H}\,s_i
\;=\;
H
\;.
\end{equation}
The TRS is called odd if $i=2$. Otherwise, the TRS is called even. The particle-hole symmetry (PHS) is implemented by Pauli matrices $\tau_i$:
\begin{equation}
\label{eq-PGS}
(\tau_i)^*\,\overline{H}\,\tau_i
\;=\;
-\,H
\;.
\end{equation}
Again it is called odd if $i=2$ and even otherwise. For the Fermi projection $P$ this implies
$$
(\tau_i)^*\,\overline{P}\,\tau_i
\;=\;
\one\,-\,P
\;.
$$ 
The natural form of PHS stemming from fermionic quadratic many-body Hamiltonians is even and implemented by $\tau_1$. This leads to the (even) Bogoliubov-de Gennes (BdG) operators:
\begin{equation}
\label{eq-BdG}
(\tau_1)^*\,\overline{H}\,\tau_1
\;=\;
-\,H
\qquad
\Longleftrightarrow
\qquad
H
\;=\;
\begin{pmatrix}
h & \Delta \\ \Delta^* & -\overline{h}
\end{pmatrix}
\;,
\end{equation}
with the so-called pair creation operator $\Delta$ satisfying the BdG equation $\Delta^T=-\Delta$ (see \cite{AZ,DS3} for details). 
Another representation of the Hamiltonian is obtained by Cayley transform on the particle-hole fiber:
\begin{equation}
\label{eq-HMaj}
H_\Maj
\;=\;
(\tau_C)^*\,H\,\tau_C
\;,
\qquad
\tau_C
\;=\;
\tfrac{1}{\sqrt{2}}
\begin{pmatrix}
\one & \imath\,\one \\ \one & -\,\imath\,\one
\end{pmatrix}
\;.
\end{equation}
Then $H_\Maj=-\overline{H_\Maj}=-(H_\Maj)^T$ is a purely imaginary and anti-symmetric operator.  Of course, it is also possible to combine PHS with TRS which leads to the 8 real CAZ classes, see Table~1. If a Hamiltonian has both a PHS and a TRS, they can be combined to a chiral symmetry $(s_i\tau_j)^*H (s_i\tau_j)=-H$. Hence the Fermi projection is described by a Fermi unitary in these cases, but this requires diagonalizing the chiral symmetry operator $s_i\tau_j$. Furthermore, the Fermi unitary then has a further symmetry property (symmetric, real, {\it etc.}). Let us regroup all these informations in a small table. The index $j$ corresponds to that of Table~1, and $s_i$ and $\sigma_i$ square to $\one$, namely $i=0,1,3$.
\begin{equation}
\label{tab-HamSym}
\begin{tabular}{|c|c|c|c|c|c|c|c|}
\hline
$j=0$ & $j=1$ & $j=2$ & $j=3$ & $j=4$ & $j=5$ & $j=6$ & $j=7$  
\\
\hline
 &  ${\scriptstyle \tau_1^*\overline{H}\tau_1=-H}$ & ${\scriptstyle\tau_1^*\overline{H}\tau_1=-H}$ & ${\scriptstyle\tau_1^*\overline{H}\tau_1=-H}$ &  & ${\scriptstyle \tau_2^*\overline{H}\tau_2=-H}$ & ${\scriptstyle \tau_2^*\overline{H}\tau_2=-H}$ & ${\scriptstyle \tau_2^*\overline{H}\tau_2=-H}$
\\ 
 ${\scriptstyle s_i^*\overline{H}s_i=H}$ & ${\scriptstyle s_i^*\overline{H}s_i=H}$ &  & ${\scriptstyle s_2^* \overline{H}s_2=H}$ & ${\scriptstyle s_2^*\overline{H}s_2=H}$ & ${\scriptstyle s_2^*\overline{H} s_2=H}$ &  & ${\scriptstyle s_i^*\overline{H} s_i=H}$
\\
\hline
 ${\scriptstyle s_i^*\overline{P}s_i=P}$ & ${\scriptstyle\sigma_i^*\overline{U}\sigma_i=U}$ & ${\scriptstyle\tau_1^*\overline{P}\tau_1=\one-P}$ & ${\scriptstyle\sigma_2^* U^T\sigma_2=U}$ & ${\scriptstyle s_2^*\overline{P}s_2=P}$ & ${\scriptstyle\sigma_2^*\overline{U}\sigma_2=U}$ & ${\scriptstyle \tau_2^*\overline{P}\tau_2=\one-P}$ & ${\scriptstyle\sigma_i^*U^T\sigma_i=U}$
\\
\hline
\end{tabular}
\end{equation}
Here the Pauli matrices $\sigma_1,\sigma_2,\sigma_3$ appear after reducing out the two symmetries in the corresponding column. These symmetry properties do, of course, have an important impact on the index pairing, in particular, when combined with similar symmetry properties of $E$ and $F$ stemming from $D$. Before starting with this analysis in Section~\ref{sec-DiracSym}, let us first discuss further symmetries of even BdG operators. The charge conservation symmetry 
\begin{equation}
\label{eq-ChargeCons}
(\tau_3)^*\,H\,\tau_3
\;=\;
H
\end{equation}
is equivalent to having a vanishing pair creation $\Delta=0$. Then $H$ is classified by $h$ which can be in CAZ class A, or AI or AII pending on whether $H$ also has a TRS. As explained in \cite{AZ} (see also \cite{DS3}), a global spin rotation invariance for a BdG operator leads to operators with odd PHS. Finally let us note that operators with PHS and/or TRS can also have another chiral symmetry or conservation law which are then implemented by further Pauli matrices (commuting with $\tau_i$ and $s_i$ as stated above).

\subsection{Symmetries of the Dirac operator}
\label{sec-DiracSym}

The Dirac operator is still given by \eqref{eq-DiracOp} with a representation $\gamma_1,\ldots,\gamma_d$ of the complex Clifford algebra. In particular, there is an associated Hardy projection $E$ which in even dimension $d$ is chiral and described by the Dirac phase $F$, see \eqref{eq-DiracPhase}. On top of that, the Clifford representation can be chosen such that $\gamma_{2i+1}=\overline{\gamma_{2i+1}}$ is real and $\gamma_{2i}=-\overline{\gamma_{2i}}$ is purely imaginary for all $i$. Then it is possible to construct for every $d$ two commuting representations $\Gamma_1,\Gamma_2,\Gamma_3$ and $\Sigma_1,\Sigma_2,\Sigma_3$ of the Pauli matrices acting on the Clifford representation space such that the following holds:
\begin{equation}
\label{tab-DiracSym}
\begin{tabular}{|c|c|c|c|c|c|c|c|}
\hline
$d=1$ & $d=2$ & $d=3$ & $d=4$ & $d=5$ & $d=6$ & $d=7$ & $d=8$ 
\\
\hline
& ${\scriptstyle\Gamma_2^*\overline{D}\Gamma_2=-D}$ & ${\scriptstyle\Gamma_2^*\overline{D}\Gamma_2=-D}$ & ${\scriptstyle\Gamma_2^*\overline{D}\Gamma_2=-D}$ &  & ${\scriptstyle\Gamma_1^*\overline{D}\Gamma_1=-D}$ & ${\scriptstyle\Gamma_1^*\overline{D}\Gamma_1=-D}$ & ${\scriptstyle\Gamma_1^*\overline{D}\Gamma_1=-D}$
\\
 ${\scriptstyle\Sigma_i^*\overline{D}\Sigma_i=D}$  & ${\scriptstyle\Sigma_i^*\overline{D}\Sigma_i=D}$ &  & ${\scriptstyle\Sigma_2^*\overline{D}\Sigma_2=D}$ & ${\scriptstyle\Sigma_2^*\overline{D}\Sigma_2=D}$ & ${\scriptstyle\Sigma_2^*\overline{D}\Sigma_2=D}$ &  & ${\scriptstyle\Sigma_i^*\overline{D}\Sigma_i=D}$
\\
\hline
 ${\scriptstyle\Sigma^*\overline{E}\Sigma=E}$ & ${\scriptstyle\Sigma^*F^T\Sigma=F}$ & ${\scriptstyle\Sigma^*\overline{E}\Sigma=\one-E}$ & ${\scriptstyle\Sigma^* \overline{F}\Sigma=F}$ & ${\scriptstyle\Sigma^*\overline{E}\Sigma=E}$ & ${\scriptstyle\Sigma^*F^T\Sigma=F}$ & ${\scriptstyle\Sigma^*\overline{E}\Sigma=\one-E}$ & ${\scriptstyle\Sigma^*\overline{F}\Sigma=F}$
\\
 & ${\scriptstyle\Sigma^*D_0^T\Sigma=D_0}$ &  & ${\scriptstyle\Sigma^* \overline{D_0}\Sigma=D_0}$ &  & ${\scriptstyle\Sigma^*D_0^T\Sigma=D_0}$ &  & ${\scriptstyle\Sigma^*\overline{D_0}\Sigma=D_0}$
\\
${\scriptstyle\Sigma^2=\one}$  & ${\scriptstyle\Sigma^2=\one}$ & ${\scriptstyle\Sigma^2=-\one}$ & ${\scriptstyle\Sigma^2=-\one}$  & ${\scriptstyle\Sigma^2=-\one}$ & ${\scriptstyle\Sigma^2=-\one}$ & ${\scriptstyle\Sigma^2=\one}$ &${\scriptstyle\Sigma^2=\one}$
\\
\hline
\end{tabular}
\end{equation}
Here $\Sigma_i$ can be $\Sigma_0=\one$, $\Sigma_1$ or $\Sigma_3$, and in the case $\Sigma_0=\one$ no matrix degree of freedom may be needed so that $\Sigma_0=1$ can be scalar. The first row is not explicitly spelled out in \cite{GS}, but can readily be obtained from the results of Section~3.4 in \cite{GS}, by applying in two of the 8 cases a Cayley transform to obtain commuting symmetry operators. Let us be explicit about this point for $d=2$. The standard representation of the Dirac opertor is, {\it cf.} Section~\ref{sec-TopInv},
$$
D
\;=\;
\begin{pmatrix}
0 & X_1 -\imath \,X_2 \\ X_1 +\imath \,X_2 & 0
\end{pmatrix}
\;,
\qquad
D\,=\,-\gamma_3\, D\,\gamma_3\;=\;\gamma_1\,\overline{D}\,\gamma_1
\;,
$$
with $\gamma_1,\gamma_2,\gamma_3$ being the Pauli matrices. From this one can read off the second line in $d=2$, namely $F={(}X_1 -\imath \,X_2{)|X_1 -\imath \,X_2|^{-1}}$ and $\Sigma=1$ scalar. Then with the Cayley transform $\gamma_C$ defined as in \eqref{eq-HMaj}, the transformed Dirac operator $D'=(\gamma_C)^*D\gamma_C$ is given by
$$
D'
\;=\;
\begin{pmatrix}
X_1 & -X_2 \\ -X_2 & -X_1
\end{pmatrix}
\;,
\qquad
D'\,=\,-(\gamma_2)^* D'\gamma_2\;=\;\overline{D'}
\;.
$$
Hence the equations in the $d=2$ entry in the first line hold for $D'$ with $\Gamma_2=\gamma_2$ and $\Sigma_i=\one$. The other case $d=6$ can be dealt with similarly. The last line of \eqref{tab-DiracSym} contains the symmetry of the Hardy projection and Dirac phase, obtained after reducing out. Here $\Sigma$ is a real symmetry on the Clifford representation space (or half of it), namely $\overline{\Sigma}=\Sigma$ and $\Sigma^2=\pm\one$. Let us stress that the tables \eqref{tab-DiracSym} and \eqref{tab-HamSym} are essentially the same, if one identifies $d\cong 1-j \mbox{ mod }8$.

\subsection{Symmetries of index pairings and $\ZM_2$-index}
\label{sec-Z2Index}

As in Section~\ref{sec-TopInv}, one can now build from the Fermi projection or Fermi unitary (stemming from the Hamiltonian) and the Hardy projection or the Dirac phase (stemming from the Dirac operator) an index pairing. Taking into account all the possible real symmetries of the Hamiltonian in \eqref{tab-HamSym} and the Dirac operator in \eqref{tab-DiracSym}, this leads to 64 index pairings, each leading to a Fredholm operator which inherits symmetry properties from \eqref{tab-HamSym} and \eqref{tab-DiracSym}. The complex pairings \eqref{eq-ComplexPairingDef} allow to pair unitaries with projections and this covers all the $\ZM$ and $2\,\ZM$ entries in Table~1. Following \cite{GS}, the pairing of two projections $P$ and $E$ is given by 
\begin{equation}
\label{eq-ProjProj}
T
\;=\;
E\,(\one-2P)\,E\,+\,\one-E
\;,
\end{equation}
the pairing of two unitaries $U$ and $F$ by 
\begin{equation}
\label{eq-UnitaryUnitary}
T
\;=\;
\frac{1}{2}\begin{pmatrix} \one & U \\ U^* & \one \end{pmatrix}
\begin{pmatrix} F & 0 \\ 0 & F \end{pmatrix}
\frac{1}{2}\begin{pmatrix} \one & U \\ U^* & \one \end{pmatrix}
\;+\;
\one_2\,-\,
\frac{1}{2}\begin{pmatrix} \one & U \\ U^* & \one \end{pmatrix}
\;.
\end{equation}
Due to the locality estimates \eqref{eq-Decay}, these are also Fredholm operators. Moreover, these index pairings as well as those given in \eqref{eq-ComplexPairingDef} inherit symmetries. These symmetries may imply that the index $\Ind(T)$ is even or vanishes. In the latter case, a secondary $\ZM_2$-valued index
$$
\Ind_2(T)
\;=\;
\dim\big(\Ker(T)\big)\;\mbox{mod }2\;\in\;\ZM_2
$$
may be well-defined. In \eqref{eq-ProjProj} and \eqref{eq-UnitaryUnitary}, one can exchange the roles of $E$ with $P$ and $U$ with $F$ respectively, without changing the numerical index $\Ind_2(T)$ \cite{GS}.

\begin{theo}[Theorem 2 in \cite{GS}]
\label{theo-indexlist}
With $P$ and $U$ having the symmetries from \eqref{tab-HamSym} and the dimension $d$ leading to the symmetries \eqref{tab-DiracSym}, the index pairings given by \eqref{eq-ComplexPairingDef}, \eqref{eq-ProjProj} and \eqref{eq-UnitaryUnitary} are well-defined and take values as stated in {\rm Table~1}.
\end{theo}

Let us stress that all these $\ZM_2$-indices are generalizations of the standard $\ZM_2$-indices used in the literature (like the Kane-Mele index for CAZ class AII for $d=2$).

\subsection{Symmetries of the spectral localizer}
\label{sec-SpecLocSym}

For the pairings of unitaries with projections the spectral localizer is defined by \eqref{eq-OddSpecLoc} and \eqref{eq-EvenSpecLoc}, respectively for even or odd pairings. Then the symmetries \eqref{tab-HamSym} and \eqref{tab-DiracSym} lead to symmetries of the spectral localizer which in turn allow to show how $\ZM$- or $\ZM_2$-valued invariants can be extracted from the finite volume spectral localizer. For the odd index pairings, this is discussed in Section 1.7 of \cite{LS1}. The proof that these invariants are indeed equal the entries of Table~1 is not given in \cite{LS1}. This as well as an exhaustive treatment of the other 32 cases involving the pairings \eqref{eq-ProjProj} and \eqref{eq-UnitaryUnitary}  is the object of another publication. The main objective here is to exhibit approximate laws that allow to find $\ZM$-valued invariants which determine the $\ZM_2$-invariants of Table~1.

\section{Integer-valued invariants for real CAZ classes}
\label{sec-RealCAZ}

In each column and line of the real CAZ classes of Table~1 there is one $\ZM$-valued invariant and one $2\,\ZM$-valued invariant. Both can be computed using the spectral localizer from the complex theory as described in Section~\ref{sec-SpecLoc}. {For reasons explained in the introductory Section~\ref{sec-Intro}, the focus is here} on the $\ZM_2$-entries of Table~1 and, more specifically, to present an approximate law for each case that allows to introduce a $\ZM$-valued invariant that modulo $2$ determines the $\ZM_2$-invariant described in Section~\ref{sec-Z2Index}.  {For periodic systems and a situation with exact symmetries or conservation laws, these additional laws correspond to those found in \cite{SS}. More precisely, for the upper and lower line of $\ZM_2$-indices in Table~1 this corresponds respectively to the case $t=1$ and $t=2$ in \cite{SS} (with $D=0$ and $\delta_\parallel=0$). Here we also prove the connection with the $\ZM_2$-index.} Furthermore, for each case a twisted spectral localizer is presented that allows to compute the new $\ZM$-valued invariant. In principle, the approach is algebraically the same as in the complex cases described in Section~\ref{sec-ComplexCAZ}. However, we will attempt to argue in each case that the approximate law can be a physically reasonable assumption. In the subsections below, we will go through all cases by following the staircase of $\ZM_2$-invariants in Table~1 downward. To give the reader some traction, we start out with zero-dimensional systems of CAZ class BDI and $D$, then follow the one-dimensional systems, {\it etc.}, up to $d=8$. The proof that the $\ZM$-invariant gives modulo $2$ the $\ZM_2$-invariant is carried out in detail only in the low-dimensional cases $d=1,2,3$, as the others are very similar.

\subsection{Zero-dimensional invariants: cases $(j,d)=(1,0)$ and $(j,d)=(2,0)$}
\label{sec-ZeroDim}

As a warm-up, let us begin with the zero-dimensional case which appears in Table~1 as the $d=8\,\mbox{mod }8$ column. The Hamiltonian is a finite dimensional matrix in this case as the physical space consists of merely one point. According to Table~1, there are two $\ZM_2$-invariants corresponding to $j=1$ and $j=2$. In both cases, the Hamiltonian is of the BdG form \eqref{eq-BdG}. The $\ZM_2$-invariant is given by the sign of the Pfaffian of $\imath$ times the Majorana representation $H_\Maj$ of the Hamiltonian, as given in \eqref{eq-HMaj}. If now the Hamiltonian has an approximate charge conservation, this $\ZM_2$-invariant can be calculated as the parity of an integer valued signature. The extra symmetry for $j=1$ w.r.t. $j=2$ is irrelevant for the following.

\begin{proposi}
Suppose that  $\eta=\|[H,\tau_3]\|$ satisfies $\eta< 2\|H^{-1}\|^{-1}$ and let $h\in\CM^{L\times L}$ be the upper diagonal entry of the BdG Hamiltonian $H\in\CM^{2L\times 2L}$, see \eqref{eq-BdG}. Then
$$
\sgn\big(\mbox{\rm Pf}
(\imath\,H_\Maj)
\big)
\;=\;
(-1)^{\frac{L^2+\Sig(h)}{2}}
\;.
$$
\end{proposi}

\noindent{\bf Proof:} First of all, the bound on $\eta$ implies that
$$
t \in[0,1] 
\;\mapsto \;
H(t)\;=\;\begin{pmatrix}
h & t\Delta\\ t\Delta^* & -\overline{h}
\end{pmatrix}
$$
is a path within the invertible matrices along which hence both sides of the equality remain unchanged. Thus it is sufficient to consider the case $\eta=0$ so that $\Delta=0$ and $H=\diag(h,-\overline{h})$ with $h$ invertible. Let $h=udu^*$ be the spectral decomposition of $h$, namely $d$ is diagonal and $u$ is unitary. There is a path $t \in[0,1] \mapsto u(t)$ of unitaries connecting $u(0)=u$ to $u(1)=\one$. Then $t \in[0,1] \mapsto \imath\,(\tau_C)^* \diag(u(t)du(t)^*,-\overline{u(t)du(t)^*})\tau_C$ is a path of real skew-adjoint invertibles connecting $\imath\,(\tau_C)^* H\tau_C$ to $\imath\,(\tau_C)^* \diag(d,-d)\tau_C$. Therefore
\begin{align*}
\sgn\big(\mbox{\rm Pf}
(\imath\,(\tau_C)^* H\tau_C)
\big)
&\;=\;
\sgn\big(\mbox{\rm Pf}
(\imath\,(\tau_C)^* \diag(d,-d)\tau_C)
\big)
\;=\;
\sgn\left(\mbox{\rm Pf}
\begin{pmatrix}
0 & -d \\ d & 0
\end{pmatrix}
\right)\\
&
\;=\;
(-1)^{\frac{L(L-1)}{2}}\sgn(\det(-d))
\;=\;
(-1)^{\frac{L(L-1)}{2}}(-1)^{\mbox{\tiny\rm Tr}(\one-p)}\;,
\end{align*}
where $p=\chi(h<0)$. From this and $\Tr(\one-p)=\frac{\Sig(h)+L}{2}$ the claim now follows.
\hfill$\Box$

\subsection{Approximate chiral symmetry: case $(j,d)=(2,1)$}

This section deals with one-dimensional BdG operators of CAZ class D. This includes a Kitaev chain without TRS. Hence $H=-(\tau_1)^*\overline{H}\tau_1$ is of the BdG form \eqref{eq-BdG}. Moreover, the Hamiltonian is supposed to have a supplementary approximate chiral symmetry implemented by a Pauli matrix $\nu_3$ commuting with the $\tau_i$, namely $\eta=\|\nu_3\,H\,\nu_3\,+\,H\|$ is required to be small. Thus one is in the setting of Section~\ref{sec-ApproxChiral} and, provided that $\eta<2g$, one has an integer valued invariant given by \eqref{eq-IndSigMod}. In particular, the spectral localizer is given by \eqref{SpecLocOddMod} and Proposition~\ref{prop-SpecLocApproxChiral} applies for $\eta<\tfrac{2g}{3}$ and admissible $(\kappa,\rho)$. This invariant then allows to compute the $\ZM_2$-invariant:

\begin{proposi}
In the above situation,
$$
\Ind_2\big(E(\one-2P)E\,+\,\one-E\big)
\;=\;
\Ind\big(E\,A|A|^{-1}\,E\,+\,\one-E\big)\;\mbox{\rm mod}\,2
\;.
$$
\end{proposi}

\noindent{\bf Proof.} For $H$ of the form \eqref{eq-HEntries}, let us consider the path of BdG Hamiltonians
$$
t \in [0,1] 
\;\mapsto\; H(t)\;=\;\begin{pmatrix}
t\,H_+ & A \\ A^* & t\,H_-
\end{pmatrix}\;.
$$
This is written in the grading of the $\nu_i$. In particular, one has $(\tau_1)^*\overline{A}\,\tau_1=-A$.
If $\eta<2g$, this is a path of invertibles so that $t \in [0,1] \mapsto P(t)=\chi(H(t)<0)$ is continuous. Moreover,  and as $(\tau_1)^*\overline{H(t)}\tau_1=-H(t)$,
$$
t \in [0,1]\; \mapsto\; 
\Ind_2\big(E(\one-2P(t))E\,+\,\one-E\big)
$$
is constant by Proposition 7 in \cite{GS}. Thus it is sufficient to show
$$
\Ind_2\big(E(\one-2P(0))E\,+\,\one-E\big)
\;=\;
\Ind\big(E\,A|A|^{-1}\,E\,+\,\one-E\big)
\,\mbox{\rm mod}\,2\;.
$$
Let $U=A|A|^{-1}$ denote the unitary phase of $A$.  Modulo 2, one has 
\begin{align*}
\Ind_2\big(E(\one-2P(0))E\,+\,\one-E\big)\;
&=\;
\dim\left(\Ker
\left(E\begin{pmatrix} 0 & U \\ U^* & 0 \end{pmatrix}E+\one-E\right)\right)
\\
&=\;
\dim\big(\Ker(EUE+\one-E)\big)\,+\,\dim\big(\Ker(EU^*E+\one-E)\big)\\
&=\;\dim\big(\Ker(EUE+\one-E)\big)\,-\,\dim\big(\Ker(EU^*E+\one-E)\big)\\
&=\;\Ind\big(EUE+\one-E\big)
\;,
\end{align*}
which concludes the proof.
\hfill$\Box$

\subsection{Approximate spin conservation: case $(j,d)=(3,1)$}
\label{sec-3,1}

Here  $j=3$ so that the Hamiltonian $H$ lies in the CAZ class DIII, namely 
\begin{equation}
\label{eq-d1DIII2}
(\tau_1)^* \,\overline{H}\,\tau_1
\;=\;
-H\;,
\qquad
(s_2)^* \,\overline{H}\,s_2
\;=\;
H
\;.
\end{equation}
Now the $3$-component of the spin is implemented in the BdG representation by $\tau_3\otimes s_3$ (see \cite{DS3}). Therefore approximate spin conservation means that  
$$
\eta\;=\;\|[H,\tau_3\otimes s_3]\|
$$
is small. Let us bring all this data into a normal form. The chiral symmetry inherited from \eqref{eq-d1DIII2} is with $\tau_1\otimes s_2$ (which is imaginary here). The suitable basis change from $\tau_i\otimes s_j$ to tensor products $\sigma_i\otimes\nu_j$ of two new sets of Pauli matrices is
$$
M\;=\;\frac{1}{\sqrt{2}}
\begin{pmatrix}
0 & 1 & 0 & 1 \\
1 & 0 & 1 & 0 \\
-\imath & 0 & \imath & 0 \\
0 & \imath & 0 & -\imath  
\end{pmatrix}
\;,
$$
because one then checks 
$$
M^*\,\tau_1\otimes s_2\, M
\;=\;\sigma_3\otimes\one\;,
\qquad
M^*\,\tau_3\otimes s_3\,M\;=\;-\;\one\otimes \nu_3
\;,
\qquad
M^T\tau_1\otimes\one M=-\imath\sigma_2\otimes  \nu_2
\;.
$$
In this basis the Hamiltonian therefore verifies
\begin{equation}
\label{eq-j=3}
(\sigma_3)^*M^*HM\,\sigma_3
\;=\;
-\,M^*HM
\;,
\qquad
(\sigma_2\otimes  \nu_2)^*\overline{M^*\,H\,M}\,\sigma_2\otimes  \nu_2
\;=\;
-\,M^*\,H\,M
\;,
\end{equation}
and 
$$
\eta\;=\;
\|[M^*HM,\nu_3]\|
\;.
$$
In particular, $M^*HM$ is  precisely in the case of item (ii) in Section~\ref{sec-ApproxSymChiral}, with the second equation in \eqref{eq-j=3} as a supplementary property. Thus, if $\eta<2g$, there is an integer index pairing $\Ind(EA_+|A_+|^{-1}E+\one-E)$ with $A_+$ given by
\begin{equation}
\label{eq-j=3H}
M^*HM
\;=\;
\begin{pmatrix}
0 & A \\ A^* & 0 
\end{pmatrix}
\;,
\qquad
A
\;=\;
\begin{pmatrix}
A_+ & B \\ C & A_-
\end{pmatrix}
\;,
\end{equation}
in the gradings of the $\sigma_i$ and $\nu_i$ respectively. Then $\eta=\|[A,\nu_3]\|=2\,\max\{\|B\|,\|C\|\}$. There is a corresponding twisted spectral localizer given in \eqref{eq-IndexPairLocCase(ii)}. Let us also note that the second equation in \eqref{eq-j=3} implies $ (\nu_2)^* A^T \nu_2=A$ and hence $(A_+)^T=A_-$.  Therefore again the sum of the indices of $A_\pm$ vanishes.  The strong $\ZM_2$-index is given in terms of this invariant:

\begin{proposi}
In the above situation,
$$
\Ind_2\big(EUE+\one-E\big)
\;=\;
\Ind\big(EA_+|A_+|^{-1}E+\one-E\big)
\,\mbox{\rm mod}\,2\;.
$$
\end{proposi}

\noindent{\bf Proof.} As $\eta<2g$, the path
$$
t \in [0,1] \;\mapsto\; A(t)\;=\;\begin{pmatrix}
A_+ & t\,B\\
t\,C & (A_+)^T
\end{pmatrix}\;,
$$
lies in the invertible operators. Thus $t \in [0,1] \mapsto U(t)=A(t)|A(t)|^{-1}$ is continuous. As $(\nu_2)^*U(t)^T\nu_2=U(t)$ and therefore
$T(t)=EU(t)E+\one-E$ satisfies $(\nu_2)^*T(t)^T\nu_2=T(t)$, it follows that $t \in [0,1] \mapsto \Ind_2(T(t))$ is constant by Proposition 5 in \cite{GS}, see also \cite{SB}. To complete the argument let us show that
$$
\Ind_2\big(E U(0)E\,+\,\one-E\big)
\;=\;
\,\Ind\big(EU_+E+\one-E\big)\;\mbox{\rm mod}\,2\;,
$$
where $U_+=A_+|A_+|^{-1}$ denotes the unitary phase of $A_+$. Indeed, modulo 2 one has 
\begin{align*}
\Ind_2\big(E U(0)E\,+\,\one-E\big)
&=\;\dim\big(\Ker(EU_+E+\one-E)\big)\,+\,\dim\big(\Ker(EU_+^TE+\one-E)\big)\\
&=\;\dim\big(\Ker(EU_+E+\one-E)\big)\,+\,\dim\big(\Ker(EU_+^*E+\one-E)\big)\\
&=\;\dim\big(\Ker(EU_+E+\one-E)\big)\,-\,\dim\big(\Ker(EU_+^*E+\one-E)\big)\\
&=\;\Ind\big(EU_+E+\one-E\big)
\;,
\end{align*}
completing the proof.
\hfill$\Box$

\subsection{Approximate conservation law: case $(j,d)=(3,2)$}
\label{sec-3,2}

The symmetry is the same as in the last Section~\ref{sec-3,1}.  Moreover, $H$ is supposed to have an approximate conservation law expressed in terms of
$$
\eta\;=\;\|[H,\tau_2]\|
\;.
$$
The observable $\tau_2$ of this approximate conservation law anticommutes with the symmetry $\tau_1\otimes s_2$ implementing the chiral symmetry. To bring this in a normal form, let us again apply the basis change $M$ from Section~\ref{sec-3,1} so that \eqref{eq-j=3} holds and $M^*HM$ is off-diagonal as in \eqref{eq-j=3H} with offdiagonal entry $A$ satisfying $ (\nu_2)^* A^T \nu_2=A$. Moreover, due to $M^*\tau_2\otimes \one M=\sigma_2\otimes \nu_2$, one has
\begin{equation}
\label{eq-j3d2}
\eta\;=\;
\|[M^*HM,\sigma_2\otimes \nu_2]\|
\;=\;
\|(\imath \nu_2 A)^*-\imath \nu_2 A \|
\;=\;
\|A+\overline{A}\|
\;.
\end{equation}
Hence this case of anticommuting (approximate) laws was already dealt with for the complex CAZ classes in Section~\ref{sec-ApproxExchange}, with a supplementary property given by the second equation in \eqref{eq-j=3}. Let us perform another basis change 
so that $(MN)^*H(MN)$ satisfies \eqref{eq-ChiralApproxEx}. This is attained by
$$
N\;=\;\frac{1}{\sqrt{2}}\begin{pmatrix}
1 & 0 & 1 & 0 \\
0 & 1 & 0 & 1 \\
0 & 1 & 0 & -1 \\
-1 & 0 & 1 & 0  
\end{pmatrix}\;,
$$ 
because indeed
$$
N^*\sigma_3\otimes \one N\;=\;\sigma_1\otimes \one\;,
\qquad
N^*\sigma_2\otimes\nu_2 N=\sigma_3\otimes \one\;,
\qquad
N^T\sigma_2\otimes\nu_2 N=\sigma_3\otimes \one\;.
$$
As in Section \ref{sec-ApproxExchange}, the diagonal entry $H_+$ of $(MN)^*H(MN)$  is invertible and for $P_+=\chi(H_+<0)$  there is a $\ZM$-valued invariant $\Ind(P_+FP_++\one-P_+)$. For $\eta<\tfrac{2g}{3}$ and $(\kappa,\rho)$ admissible, it can be calculated via the spectral localizer given in  \eqref{eq-ChiralApproxExSpecLoc}. This invariant allows to compute the $\ZM_2$-invariant.

\begin{proposi}
In the above set-up,
$$
\Ind_2\big(EUE+\one-E\big)
\;=\;
\Ind\big(P_+FP_++\one-P_+\big)\,\mbox{\rm mod}\,2\;.
$$
\end{proposi}

\noindent{\bf Proof.}
Carrying out the basis change $N$, one directly checks $H_+=\imath\tfrac{1}{2}(A-\overline{A})\nu_2$. As $\eta<2g$, the path $t \in [0,1] \mapsto A(t) = A-\tfrac{1}{2}t(A+\overline{A})$ is within the invertibles. Thus $t \in [0,1] \mapsto U(t)=A(t)|A(t)|^{-1}$ is continous. Moreover, $(\nu_2)^* U(t)^T \nu_2=U(t)$ so that
$$
t \in [0,1]\; \mapsto\; \Ind_2\big(EU(t)E+\one-E\big)
$$
is constant by Proposition 5 in \cite{GS}, see also \cite{SB}. Thus it is sufficient to show
$$
\Ind_2\big(EU(1)E+\one-E\big)
\;=\;
\Ind\big(P_+FP_++\one-P_+\big)\,\mbox{\rm mod}\,2\;.
$$
As 
$$
\one-2P_+
\;=\;
\imath\tfrac{1}{2}(A-\overline{A})\nu_2|\imath\tfrac{1}{2}(A-\overline{A})\nu_2|^{-1}
\;=\;
\imath\tfrac{1}{2}(A-\overline{A})|\imath\tfrac{1}{2}(A-\overline{A})|^{-1}\nu_2
\;=\;
\imath\, U(1)\nu_2
\;,
$$
one has
\begin{align*}
\Ind_2\big(E U(1)E+\one-E\big)
&\;=\;
\Ind_2\big(E\,\imath \,U(1)\nu_2E+\one-E\big)\\
&\;=\;
\dim\big(\Ker((\one-2P_+)F+F(\one-2P_+))\big)\,\mbox{\rm mod}\,2
\\
&\;=\;\dim\big(\Ker(F^*(\one-2P_+)F+(\one-2P_+))\big) \,\mbox{\rm mod}\,2
\\
&
\;=\;\Ind\big(P_+FP_++\one-P_+\big)\,\mbox{\rm mod}\,2
\;,
\end{align*}
where the second equality holds by Proposition 2 in \cite{GS}. 
\hfill$\Box$

\subsection{Approximately conserved odd spin: case $(j,d)=(4,2)$}
\label{sec-SQHE}

This section deals with two-dimensional systems with odd TRS $(s_2)^*\overline{H}s_2=H$. Such systems have a $\ZM_2$-invariant, allowing to distinguish a trivial system from a quantum spin Hall system. Let us suppose that $H$ has an approximate spin conservation in the sense that $ \eta=\|[H,s_3]\|$ is small. By the results of Section~\ref{sec-ApproxExchange2}, the bound $\eta<2g=2\|H^{-1}\|^{-1}$ then allows to consider two invariants $\Ind(P_\pm FP_\pm+\one -P_\pm)$ which, moreover, satisfy \eqref{eq-SumIndex}. This actually does not pend on the odd TRS which does, on the other hand, imply $(s_2)^*\overline{P} s_2=P$ so that combined with $F^T=F$ one deduces $\Ind(P FP+\one -P)=0$. Thus the sum of the invariants $\Ind(P_\pm FP_\pm+\one -P_\pm)$ vanishes and Proposition~\ref{prop-SpecLocApproxConsLaw} implies that for $\eta < \tfrac{2g}{3}$ and admissible $(\kappa,\rho)$
$$
\Ind\big(P_+ FP_++\one -P_+\big)
\;=\;
\frac{1}{4}\;
\Sig(L_{\kappa,\rho})
\;,
$$
where $L_{\kappa,\rho}$ is given by \eqref{eq-SpecLocConsLaw}. The invariants $\Ind(P_\pm FP_\pm+\one -P_\pm)$ are called the spin Chern numbers because for covariant systems there is an index theorem showing that these invariants are equal to the Chern numbers $\Ch(P_\pm)$ \cite{PS}. They were initially introduced by Sheng {\it et al.} \cite{SWSH} for periodic systems with twisted boundary conditions and then by Prodan  \cite{Pro} in the form discussed at the end of Section~\ref{sec-ApproxExchange2}. As already stressed before, the condition $\eta<2g$ is less stringent than the one used in \cite{Pro}. The physical implication of non-vanishing Chern numbers on boundary currents is discussed in \cite{SB2}. That the parity of the spin Chern numbers is equal to the $\ZM_2$-invariant was already proved in \cite{SB}, but the following argument is more direct.

\begin{proposi}
For $\eta < 2g$,
$$
\Ind_2\big(PFP+\one-P\big)
\;=\;
\Ind\big(P_+FP_++\one-P_+\big)\,\mbox{\rm mod}\,2\;.
$$
\end{proposi}

\noindent{\bf Proof.} Recall that in the grading of $s_3$,
$$
H\;=\;
\begin{pmatrix}
H_+ & A \\ A^* & \overline{H_+}
\end{pmatrix}
\;,
\qquad
\eta\,=\,2\|A\|
\;.
$$
Moreover, $P=\chi(H\leq 0)$, $P_+=\chi(H_+<0)$ and $F=(X_1+\imath X_2)|X_1+\imath X_2|^{-1}$. By the bound on $\eta$,
$$
t \in [0,1] 
\;\mapsto\; 
H(t)\;=\;
\begin{pmatrix}
H_+ & tA \\t A^* & \overline{H_+}
\end{pmatrix}
$$
is a path of invertibles with odd TRS. Hence $t \in [0,1] \mapsto P(t)=\chi(H(t)<0)$ is continuous and  by Proposition 5 in \cite{GS}
$$
t \in [0,1] \;\mapsto\; \Ind_2\big(P(t) FP(t)+\one-P(t)\big)
$$
is constant, see also \cite{SB}. Thus the claim follows from
$$
\Ind_2\big(P(0)FP(0)+\one-P(0)\big)
\;=\;
\Ind\big(P_+FP_++\one-P_+\big)\,\mbox{\rm mod}\,2\;.
$$
As $P(0)=\diag(P_+,\overline{P_+})$, one has modulo 2 
\begin{align*}
\Ind_2&\big(P(0)FP(0)+\one-P(0)\big)\\
&=\;\dim\big(\Ker(P_+FP_++\one-P_+)\big)+\dim\big(\Ker(\overline{P_+}F\overline{P_+}+\overline{\one-P_+})\big)\\
&=\;\dim\big(\Ker(P_+FP_++\one-P_+)\big)+\dim\big(\Ker(P_+F^*P_++\one-P_+)\big)\\
&=\;\Ind\big(P_+FP_++\one-P_+\big)\;,
\end{align*}
concluding the proof.
\hfill$\Box$

\subsection{Approximate spin inversion symmetry: case $(j,d)=(4,3)$}
\label{sec-4,3}

 The symmetry is the same as in Section \ref{sec-SQHE}, namely $H$ satisfies $ s_2\overline{H} s_2=H$. Now approximate spin inversion symmetry means that 
$$
\eta\;=\;\| s_3H s_3\,+\,H\|
\;.
$$
is small. Thus one is in the setting of Section \ref{sec-ApproxChiral} and provided that $\eta<2g$, one has an integer valued invariant given by \eqref{eq-IndSigMod}. For $\eta < \tfrac{2g}{3}$ and $(\kappa, \rho)$ admissible, it can be calculated via the spectral localizer given in {\eqref{SpecLocOddMod}}. This invariant allows to compute the $\ZM_2$-invariant:

\begin{proposi}
In the above situation,
$$
\Ind_2\big(E(\one-2P)E+\one-E\big)
\;=\;
\Ind\big(EA|A|^{-1}E+\one-E\big)\,\mbox{\rm mod}\,2\;.
$$
\end{proposi}

\noindent{\bf Proof.}
For $H$ of the form \eqref{eq-HEntries}, let us consider the path of Hamiltonians
$$
t \in [0,1] \mapsto H(t)\;=\;\begin{pmatrix}
tH_+ & A \\ A^* & tH_-
\end{pmatrix}\;,
$$
in the grading of the $s_i$. In particular, one has $A=-A^T$. For $\eta<2g$ this is a path of invertibles, so $t \in [0,1] \mapsto P(t)=\chi(H(t)<0)$ is continuous. Moreover, and as $ s_2\overline{H(t)} s_2=H(t)$
$$
t \in [0,1] \mapsto \Ind_2\big(E(\one-2P(t))E+\one-E\big)
$$
is constant by Proposition 5 in \cite{GS}, see also \cite{SB}. Thus it is sufficient to show
$$
\Ind_2\big(E(\one-2P(0))E+\one-E\big)
\;=\;
\Ind\big(EA|A|^{-1}E+\one-E\big)\,\mbox{\rm mod}\,2\;.
$$
Let $U=A|A|^{-1}$ denote the unitary phase of $A$, so $\one-2P(0)=\binom{0\;\;\;U}{U^*\;0}$. Modulo 2 one has
\begin{align*}
\Ind_2&\big(E(\one-2P(0))E+\one-E\big)\\
&=\;\dim\big(\Ker(EUE+\one-E)\big)+\dim\big(\Ker(EU^*E+\one-E)\big)\\
&=\;\dim\big(\Ker(EUE+\one-E)\big)-\dim\big(\Ker(EU^*E+\one-E)\big)\\
&=\;\Ind\big(EUE+\one-E\big)\;,
\end{align*}
completing the argument.
\hfill$\Box$

\subsection{Approximate conservation law: case $(j,d)=(5,3)$}
\label{sec-j5d3}

As $j=5$, one has
\begin{equation}
\label{eq-d3j5}
(\tau_2)^* \,\overline{H}\,\tau_2
\;=\;
-H\;,
\qquad
( s_2)^* \,\overline{H}\,s_2
\;=\;
H
\;.
\end{equation}
Suppose that
$$
\eta\;=\;\|[H,\tau_1\otimes s_1]\|
$$
is small. One way to interprete this is as a particle-hole exchange togehter with a spin inversion along the $1$-direction. Another way to look at it is as an approximate conservation law that commutes with the chiral symmetry $\tau_2 \otimes s_2$. This case is covered as item (ii) in Section~\ref{sec-ApproxSymChiral}. Let us bring this into a normal form. The suitable basis change from $\tau_i\otimes s_j$ to tensor poducts $\sigma_i\otimes \nu_j$ of two new sets of Pauli matrices is
$$
M\;=\;\frac{1}{\sqrt{2}}
\begin{pmatrix}
0 & 1 & 1 & 0 \\
1 & 0 & 0 & -1 \\
1 & 0 & 0 & 1 \\
0 & -1 & 1 & 0  
\end{pmatrix}
\;.
$$
This leads to 
$$
M^*\,\tau_2\otimes  s_2\, M
\;=\;\sigma_3\otimes\one\;,
\quad 
M^*\,\tau_1\otimes s_1\,M\;=\;\one\otimes \nu_3
\;.
$$
In this basis one hence has
$$
(\sigma_3)^*M^*HM\sigma_3\;=\;-M^*HM\;,
\quad
\eta\;=\;\|[M^*HM,\nu_3]\|\;,
$$
So $M^*HM$ is in the case of item (ii) in Section~\ref{sec-ApproxSymChiral}
$$
M^*\,H\,M\;=\;
\begin{pmatrix} 0 & A \\ A^* & 0 \end{pmatrix}
\;,
\qquad
A\,=\,\begin{pmatrix} A_+ & B \\ C & A_- \end{pmatrix}
\;,
$$
in the gradings of $\sigma_i$ and $\nu_i$ respectively. The TRS in \eqref{eq-d3j5} leads, due to $M^*\one\otimes s_2 M=-\one\otimes \nu_2$, to $(\one\otimes \nu_2)^*\overline{M^*\,H\,M}\,\one\otimes \nu_2=M^*\,H\,M$. This implies $ (\nu_2)^* \overline{A} \nu_2=A$ and hence $\overline{A_+}=A_-$. The integer valued invariant and the corresponding spectral localizer are given in \eqref{eq-IndexPairLocCase(ii)}. The strong $\ZM_2$-index is given in terms of this invariant:

\begin{proposi}
In the above situation,
$$
\Ind_2\big(EUE+\one-E\big)
\;=\;
\Ind\big(EA_+|A_+|^{-1}E+\one-E\big)\,\mbox{\rm mod}\,2\;.
$$
\end{proposi}

\noindent{\bf Proof.}
As $\eta<2g$
$$
t \in [0,1] \mapsto A(t)\;=\;\begin{pmatrix}
A_+ & tB\\
tC & \overline{A_+}
\end{pmatrix}\;,
$$
lies in the invertible operators. Thus $t \in [0,1] \mapsto U(t)=A(t)|A(t)|^{-1}$ is continuous and as $ (\nu_2)^*\overline{U(t)} \nu_2=U(t)$ for all $t \in [0,1]$
$$
t \in [0,1] \mapsto \Ind_2\big(EU(t)E+\one-E\big)\;,
$$
is constant by Proposition 7 in \cite{GS}. The claim follows from
$$
\Ind_2\big(EU(0)E+\one-E\big)
\;=\;
\Ind\big(EU_+E+\one-E\big)\,\mbox{\rm mod}\,2\;,
$$
where $U_+=A_+|A_+|^{-1}$ denotes the unitary phase of $A_+$. Modulo 2 one has 
\begin{align*}
\Ind_2\big(EU(0)E+\one-E\big)
\;&=\;\dim\big(\Ker(EU_+E+\one-E)\big)+\dim\big(\Ker(E\overline{U_+}E+\one-E)\big)\\
&=\;\dim\big(\Ker(EU_+E+\one-E)\big)+\dim\big(\Ker(\overline{(\one-E)U_+(\one-E)}+\overline{E})\big)\\
&=\;\dim\big(\Ker(EU_+E+\one-E)\big)-\dim\big(\Ker(EU_+^*E+\one-E_0)\big)\\
&=\;\Ind\big(EU_+E+\one-E\big)\;,
\end{align*}
where the second step follows as $\Sigma^* \overline{E}\Sigma=\one-E$ for some odd symmetry operator $\Sigma$ commuting with $U_+$ by \eqref{tab-DiracSym} and the third step follows from Lemma 1 in \cite{GS}.
\hfill$\Box$

\vspace{.2cm}

Let us note that there is another way to attain the case $(j,d)=(5,3)$. Instead of starting with a BdG Hamiltonian as in \eqref{eq-d3j5}, one could have a system with chiral symmetry and odd TRS (for which the above basis change would then be irrelevant). The approximate conservation is then simply the $z$-component of the spin.

\subsection{Approximate charge conservation: case $(j,d)=(5,4)$}

The symmetry is the same as in Section \ref{sec-j5d3}, that is, $H$ satisfies \eqref{eq-d3j5}. Moreover, the BdG Hamiltonian $H$ is supposed to have an approximate charge conservation law expressed in terms of 
$$
\eta\;=\;\|[H,\tau_3]\|
\;.
$$
There is a basis change $M$ passing from tensor products $\tau_i\otimes s_j$ to tensor products $\sigma_i \otimes \nu_j$ of two new sets of Pauli matrices such that
$$
M^*\,\tau_2\otimes s_2\, M
\;=\;-\;\sigma_3\otimes\one\;,
\quad
M^*\tau_3\otimes \one M\;=\;\sigma_1\otimes \one\;,
$$
this leads to 
$$
(\sigma_3)^* \,M^*\,H\,M\,\sigma_3
\;=\;-\,M^*\,H\,M
\;,
\qquad
\eta
\;=\;
\|
[M^*\,H\,M,\sigma_1]
\|
\;.
$$
Hence one is in the situation of Section~\ref{sec-ApproxExchange}, that is,
$$
M^*HM\;=\;
\begin{pmatrix}
0 & A\\
A^* & 0
\end{pmatrix}
$$ 
in the grading of $\sigma_3$ with $\eta=\|A-A^*\|$. As in Section~\ref{sec-ApproxExchange}, let us conjugate by the eigenbasis $N$ of $\sigma_1$ and then set $\widehat{H}=N^*M^*HMN$. This leads to 
$$
(\sigma_1)^* \,\widehat{H}\,\sigma_1
\;=\;-\;\widehat{H}
\;,
\qquad
\eta
\;=\;
\|
[\widehat{H},\sigma_3]
\|
\;.
$$
Hence $\widehat{H}$ is of the form \eqref{eq-ChiralApproxEx2}. Moreover, $M$ can be chosen such that $M^*(\tau_2\otimes\one)M=\sigma_3\otimes\nu_2$, then the TRS in \eqref{eq-d3j5} leads to $(\sigma_3\otimes\nu_2)^* \overline{\widehat{H}} \sigma_3\otimes\nu_2=-\widehat{H}$. As in Section \ref{sec-ApproxExchange}, the diagonal entry $H_+=\tfrac{1}{2}(A^*+A)$ of $\widehat{H}$ is invertible and for $P_+=\chi(H_+<0)$  there is a $\ZM$-valued invariant $\Ind(P_+FP_++\one-P_+)$. For $\eta<\tfrac{2g}{3}$ and $(\kappa,\rho)$ admissible, it can be calculated via the spectral localizer given in  \eqref{eq-ChiralApproxExSpecLoc}. As for $(j,d)=(3,2)$, one can show that this invariant allows to compute the $\ZM_2$-invariant:
$$
\Ind_2\big(EUE+\one-E\big)
\;=\;
\Ind\big(P_+FP_++\one-P_+\big)\,\mbox{\rm mod}\,2\;.
$$
%

\subsection{Approximate charge conservation: case $(j,d)=(6,4)$}
\label{sec-j6d4}

As $j=6$,  the Hamiltonian satisfies $(\tau_2)^*\overline{H}\tau_2=-H$. Moreover, let us again suppose that charge is approximately conserved, namely $\eta=\|[H,\tau_3]\|$ is small. Thus one is in the setting of Section~\ref{sec-ApproxExchange2}. Moreover, the TRS implies $H_-=-\overline{H_+}$. For $\eta<2g$ one has an integer valued invariant given by $\Ind(P_+FP_++\one-P_+)$. For $\eta<\tfrac{2g}{3}$ one can calculate this invariant via the spectral localizer given in \eqref{eq-SpecLocConsLaw}. As for $(j,d)=(4,2)$, one can show that the strong $\ZM_2$-index is given in terms of this invariant:
$$
\Ind_2\big(PFP+\one-P\big)
\;=\;
\Ind\big(P_+FP_++\one-P_+\big)\,\mbox{\rm mod}\,2\;.
$$
%

\subsection{Approximate spin inversion: case $(j,d)=(6,5)$}

The symmetry is the same as in Section~\ref{sec-j6d4}, namely $H$ satisfies $(\tau_2)^*\overline{H}\tau_2=-H$. Now approximately spin inversion symmetry means that 
$$
\eta\;=\;\|s_3Hs_3\,+\,H\|
$$
is small. Thus one is in the setting of Section \ref{sec-ApproxChiral} and provided that $\eta<2g$, one has an integer valued invariant given by \eqref{eq-IndSigMod}. For $\eta < \tfrac{2g}{3}$ and $(\kappa, \rho)$ admissible, it can be calculated via the spectral localizer given in {\eqref{SpecLocOddMod}}. As for $(j,d)=(2,1)$, one can show that this invariant allows to compute the $\ZM_2$-invariant:
$$
\Ind_2\big(E(\one-2P)E+\one-E\big)
\;=\;
\Ind\big(EA|A|^{-1}E+\one-E\big)\,\mbox{\rm mod}\,2\;.
$$
%

\subsection{Approximate chiral symmetry: case $(j,d)=(7,5)$}
\label{sec-OddPHS}

After reducing out the SU$(2)$ invariance of a BdG-Hamiltonian with spin $\frac{1}{2}$ (see \cite{AZ,DS3}) one gets
$$
H
\;=\;
\begin{pmatrix}
h & 0 & 0 & \Delta \\ 
0 & h & -\Delta & 0 \\
0 & - \overline{\Delta} & -\overline{h} & 0 \\
\overline{\Delta} & 0 & 0 & -\overline{h} 
\end{pmatrix}
\;,
$$
with $\Delta^T=\Delta$.
Thus one has two reduced operators
$$
H_{\mbox{\rm\tiny red}}
\;=\;
\begin{pmatrix}
h & \pm\Delta \\ \pm\overline{\Delta} &-\overline{h}
\end{pmatrix}
\;.
$$
Each has the odd PHS 
\begin{equation}
\label{eq-oddPHS}
(\tau_2)^*\overline{H_{\mbox{\rm\tiny red}}}\,\tau_2=-H_{\mbox{\rm\tiny red}}
\;.
\end{equation}
The odd TRS (due to spin $\frac{1}{2}$) is $( s_2)^*\overline{H} s_2=H$. Writing this out leads for the reduced operators to
\begin{equation}
\label{eq-oddPHSTRS}
\overline{H_{\mbox{\rm\tiny red}}}\;=\;H_{\mbox{\rm\tiny red}}
\;.
\end{equation}
Summing up, a class CI system is specified by \eqref{eq-oddPHS} and \eqref{eq-oddPHSTRS}. It is now useful to pass to a representation that diagonalizes the associated chiral symmetry $(\tau_2)^*H_{\mbox{\rm\tiny red}}\tau_2=-H_{\mbox{\rm\tiny red}}$. With the Cayley transform $\tau_C$ defined as in \eqref{eq-HMaj} one has $\tau_C\tau_2\tau_C^*=\tau_3$.  Thus
$$
\tau_CH_{\mbox{\rm\tiny red}}\tau_C^*
\;=\;
\begin{pmatrix}
0 & h\,\mp\,\imath\Delta \\ (h\,\mp\,\imath\Delta)^* & 0
\end{pmatrix}
\;,
$$
is indeed off-diagonal, and the off-diagonal entry $A=h\mp\imath\Delta$ satisfies $A^T=A$.  For $d=5$, this implies that $\Ind_2(EA|A|^{-1}E+\one-E)=0$ \cite{GS}. Now let us suppose that the Hamiltonian has an approximate chiral symmetry implemented in an extra sublattice degree of freedom by a Pauli matrix $\nu_3$, namely
$$
\eta\;=\;\|\nu_3 H\nu_3+H\|
$$
is small. This case is covered as item (i) in Section~\ref{sec-ApproxSymChiral}. Hence let the entries of $A$ in the grading of $\nu_i$ be given as in \eqref{eq-AEntries}, namely $A=\begin{pmatrix}
A_+ & B \\ C & A_- \end{pmatrix}$. Then $A^T=A$ implies $C=B^T$. The integer valued invariant and the correspondig spectral localizer are given in \eqref{eq-IndexPairLocCase(i)}. As for $(j,d)=(3,1)$, one can show that the strong $\ZM_2$-index is given in terms of this invariant:
$$
\Ind_2\big(EUE+\one-E\big)
\;=\;
\Ind\big(EB|B|^{-1}E+\one-E\big)\,\mbox{\rm mod}\,2\;.
$$
%

\subsection{Approximate conservation law: case $(j,d)=(7,6)$}

As again $j=7$, let us work in the framework of Section~\ref{sec-OddPHS}. We identify $H$ with {$\tau_CH_{\mbox{\rm\tiny red}}\tau_C^*$}. Moreover, we assume that there is again a sublattice degree of freedom with associated Pauli matrices $\nu_1,\nu_2,\nu_3$, and that
$$
\eta\;=\;
\|({\tau}_1\otimes \nu_2)^*H{\tau}_1\otimes \nu_2+ H\|\;=\;\|[H, {\tau}_2\otimes\nu_2]\|\;=\;\|\nu_2^*A^*\nu_2+A\|
$$
is small. Hence one can proceed similarly as in the case $(j,d)=(3,2)$. We applay the same basis change $N$ as in Section~\ref{sec-3,2}. Because 
$$
N^*{\tau}_3\otimes \one N\;=\;{\tau}_1\otimes \one\;,
\qquad
N^*{\tau}_2\otimes\nu_2 N\;=\;{\tau}_3\otimes \one\;,
\qquad
N^T{\tau}_2\otimes\one N\;=\;{\tau}_3\otimes \nu_2\;,
$$
${N^*HN}$ satisfies \eqref{eq-ChiralApproxEx}.
As in Section \ref{sec-ApproxExchange}, the diagonal entry $H_+$ of ${N^*HN}$  is invertible and for $P_+=\chi(H_+<0)$  there is a $\ZM$-valued invariant $\Ind(P_+FP_++\one-P_+)$. For $\eta<\tfrac{2g}{3}$ and $(\kappa,\rho)$ admissible, it can be calculated via the spectral localizer given in  \eqref{eq-ChiralApproxExSpecLoc}. As for $(j,d)=(3,2)$, one can show that this invariant allows to compute the $\ZM_2$-invariant:
$$
\Ind_2\big(EUE+\one-E\big)
\;=\;
\Ind\big(P_+FP_++\one-P_+\big)\,\mbox{\rm mod}\,2\;.
$$
%

\subsection{Approximately conserved even spin: case $(j,d)=(0,6)$}
\label{sec-j0d6}

As $j=0$, the Hamiltonian fulfills an even TRS which is supposed to be of the form $s_1 \overline{H} s_1=H$. This may not come directly from spinless particles, but after reducing out other internal degrees of freedom. Then suppose that
$$
\eta
\;=\;
\|[H,s_3]\|
$$
is small. Thus one is in the setting of Section~\ref{sec-ApproxExchange2}. Moreover, the TRS implies $H_-=\overline{H_+}$. For $\eta<2g$, one has an integer valued invariant given by $\Ind(P_+FP_++\one-P_+)$. For $\eta<\tfrac{2g}{3}$ and $(\kappa, \rho)$ admissible, one can calculate this invariant via the spectral localizer given in \eqref{eq-SpecLocConsLaw}. As for $(j,d)=(4,2)$, one can show that the strong $\ZM_2$-index is given in terms of this invariant:
$$
\Ind_2\big(PFP+\one-P\big)
\;=\;
\Ind\big(P_+FP_++\one-P_+\big)\,\mbox{\rm mod}\,2\;.
$$
%

\subsection{Approximate spin inversion symmetry: case $(j,d)=(0,7)$}

The symmetry is the same as in Section \ref{sec-j0d6}, namely $s_1 \overline{H} s_1=H$. Then suppose that
$$
\eta
\;=\;
\|s_3Hs_3\,+\,H\|
$$
is small. Thus one is in the setting of Section~\ref{sec-ApproxChiral}. Moreover, the TRS implies $A=A^T$. Provided that $\eta<2g$, one has an integer valued invariant given by \eqref{eq-IndSigMod}. For $\eta < \tfrac{2g}{3}$ and $(\kappa, \rho)$ admissible, it can be calculated via the spectral localizer given in {\eqref{SpecLocOddMod}}. As for $(j,d)=(2,1)$ one can show that this invariant allows to compute the $\ZM_2$-invariant:
$$
\Ind_2\big(E(\one-2P)E+\one-E\big)
\;=\;
\Ind\big(EA|A|^{-1}E+\one-E\big)\,\mbox{\rm mod}\,2\;.
$$
%

\subsection{Approximate conservation law: case $(j,d)=(1,7)$}

 As $j=1$, one has  
\begin{equation}
\label{eq-j1d7}
(\tau_1)^* \,\overline{H}\,\tau_1
\;=\;
-H\;,
\qquad
(s_1)^* \,\overline{H}\,s_1
\;=\;
H
\;.
\end{equation}
Suppose that 
$$
\eta\;=\;\|[H,\tau_2\otimes s_2]\|
$$
is small. Let us bring this into a normal form. The chiral symmetry inherited from \eqref{eq-j1d7} is with $\tau_1\otimes s_1$. There is a basis change from $\tau_i\otimes s_j$ to tensor products $\sigma_i\otimes \nu_j$ of two new sets of Pauli matices such that
$$
M^*\,\tau_1\otimes s_1\, M
\;=\;\sigma_3\otimes\one\;,
\quad
M^*\,\tau_2\otimes s_2\,M\;=\;-\one\otimes\nu_3
\;.
$$
In this basis one hence has
$$
(\sigma_3)^*M^*HM\sigma_3
\;=\;-M^*HM\;,
\quad
\eta\;=\;\|[M^*HM, \nu_3]\|\;.
$$
So $M^*HM$ is in the case of item (ii) in Section~\ref{sec-ApproxSymChiral}
$$
M^*\,H\,M\;=\;
\begin{pmatrix} 0 & A \\ A^* & 0 \end{pmatrix}
\;,
\qquad
A\,=\,\begin{pmatrix} A_+ & B \\ C & A_- \end{pmatrix}
\;,
$$
in the gradings of $\sigma_i$ and $\nu_i$ respectively. Moreover the basis change $M$ can be chosen such that  $M^*\,\one\otimes s_1\,M=\one\otimes\nu_1$, thus the TRS in \eqref{eq-j1d7} leads to 
$
(\one\otimes\nu_1)^*\overline{M^*\,H\,M}\,\one\otimes\nu_1
=
\,M^*\,H\,M
$.
This implies $ (\nu_1)^* \overline{A} \nu_1=A$ and hence $\overline{A_+}=A_-$. The integer valued invariant and the correspondig spectral localizer are given in \eqref{eq-IndexPairLocCase(ii)}. As for $(j,d)=(3,1)$ one can show that the strong $\ZM_2$-index is given in terms of this invariant:
$$
\Ind_2\big(EUE+\one-E\big)
\;=\;
\Ind\big(EA_+|A_+|^{-1}E+\one-E\big)\,\mbox{\rm mod}\,2\;.
$$
%

\subsection{Approximate conservation law: case $(j,d)=(1,8)$}

One again has an even spin, but here we have to suppose that the spin is {\em not} $0$. For sake of concreteness, let us suppose that the spin is $1$ so that a three-dimensional representation $s^1,s^2,s^3$ of su$(2)$ is relevant which we choose (as  in the appendix of \cite{DS3})  to be
$$
s^3
\;=\;
\begin{pmatrix}
1 & 0 & 0 \\ 0 & 0 & 0 \\ 0 & 0 & -1
\end{pmatrix}
\;,
\qquad
e^{\imath \pi s^2}
\;=\;
\begin{pmatrix}
0 & 0 & -1 \\ 0 & -1 & 0 \\ -1 & 0 & 0
\end{pmatrix}
\;.
$$
Let $H$ satisfy 
\begin{equation}
\label{eq-j1d8}
(\tau_1)^* \,\overline{H}\,\tau_1
\;=\;
-H\;,
\qquad
(e^{\imath \pi s^2})^* \,\overline{H}\, e^{\imath \pi s^2}
\;=\;
H
\;.
\end{equation}
Suppose that
$$
\eta\;=\;\|[H,\tau_3\otimes e^{\imath \pi s^2}]\|\;,
$$
is small. The chiral symmetry is with $\tau_1\otimes e^{\imath \pi s^2}$. There is a basis change $M$ such that 
$$
M^*\,\tau_1\otimes e^{\imath \pi s^2}\, M
\;=\;-\;\sigma_3\otimes\one\;,
\quad
M^*\tau_3\otimes e^{\imath \pi s^2}  M^*\;=\;\sigma_1\otimes \one\;.
$$
This leads to 
$$
(\sigma_3)^* \,M^*\,H\,M\,\sigma_3
\;=\;M^*\,H\,M
\;,
\qquad
\eta
\;=\;
\|
[M^*\,H\,M,\sigma_1]
\|
\;.
$$
Hence one is in the situation of Section~\ref{sec-ApproxExchange}, and thus, in the grading of $\sigma_3$,
$$
M^*HM\;=\;
\begin{pmatrix}
0 & A\\
A^* & 0
\end{pmatrix}
\;.
$$ 
As in Section~\ref{sec-ApproxExchange}, let us conjugated by the eigenbasis $N$ of $\sigma_1$ and then set $\widehat{H}=N^*M^*HMN$. This leads to 
$$
(\sigma_1)^* \,\widehat{H}\,\sigma_1
\;=\;-\;\widehat{H}
\;,
\qquad
\eta
\;=\;
\|
[\widehat{H},\sigma_3]
\|
\;.
$$
Hence $\widehat{H}$ is of the form {\eqref{eq-ChiralApproxEx2}}. One can chose $M$ such that $M^*\one\otimes e^{\imath \pi s^2}M=\one\otimes e^{\imath \pi s^2}$, so the  TRS in \eqref{eq-j1d8} leads to  $(\one\otimes e^{\imath \pi s^2})^* \overline{\widehat{H}} \one\otimes e^{\imath \pi s^2}={+}\widehat{H}$. As in Section \ref{sec-ApproxExchange}, $H_+=\tfrac{1}{2}(A^*+A)$ is invertible and for $P_+=\chi(H_+<0)$  there is a $\ZM$-valued invariant $\Ind(P_+FP_++\one-P_+)$. For $\eta<\tfrac{2g}{3}$ and $(\kappa,\rho)$ admissible, it can be calculated via the spectral localizer given in  \eqref{eq-ChiralApproxExSpecLoc}. As for $(j,d)=(3,2)$ one can show that this invariant allows to compute the $\ZM_2$-invariant:
$$
\Ind_2\big(EUE+\one-E\big)
\;=\;
\Ind\big(P_+FP_++\one-P_+\big)\,\mbox{\rm mod}\,2\;.
$$
%

\subsection{Approximately conserved charge: case $(j,d)=(2,8)$}

As $j=6$  the Hamiltonian satisfies $\tau_1\overline{H}\tau_1=-H$. Moreover charge is approximately conserved, namely $\eta=\|[H,\tau_3]\|$ is supposed to be small. This is exactly as in the zero-dimensional case discussed in Section~\ref{sec-ZeroDim}. Thus one is in the setting of Section \ref{sec-ApproxExchange2}. Moreover the TRS implies $H_-=-\overline{H_+}$. For $\eta<2g$ one has an integer valued invariant given by $\Ind(P_+FP_++\one-P_+)$. For $\eta<\tfrac{2g}{3}$ one can calculate this invariant via the spectral localizer given in \eqref{eq-SpecLocConsLaw}. As for $(j,d)=(4,2)$, one can show that the strong $\ZM_2$-index is given in terms of this invariant:
$$
\Ind_2\big(PFP+\one-P\big)
\;=\;
\Ind\big(P_+FP_++\one-P_+\big)\,\mbox{\rm mod}\,2\;.
$$
%

\appendix

\section{Proofs of properties of the twisted spectral localizers}
\label{app-SpecLocProof}

\noindent{\bf Proof of Proposition \ref{prop-SpecLocApproxChiral}.}
Let us begin by verifying that
$$
t\in[0,1]
\;\mapsto\;
L_{\kappa, \rho}(t)
\;=\;
\begin{pmatrix} \kappa\,\sigma_3D & H(t) \\ H(t) & \kappa\,\sigma_3D\end{pmatrix}_\rho
\quad
\mbox{ with }
\quad
H(t)
\;=\;
\begin{pmatrix} tH_+ & A \\ A^* & tH_-\end{pmatrix}
$$
is a path of invertibles. By the bound on $\eta$, one has $\|H-H(0)\|<\frac{g}{3}$. Thus $g(0)=\|H(0)^{-1}\|^{-1}$ satisfies
$$
\frac{2g}{3}\;<\;g(0)\;<\;\frac{4g}{3}
\;.
$$
Therefore \eqref{eq:kappa2} implies
$$
\kappa \;\leq \;\frac{2g^3}{81\|H\|\|[D,A]\|}\;<\;\frac{2(\tfrac{3}{2}g(0))^3}{81\|H(0)\|\|[D,A]\|}
\;=\;
\frac{g(0)^3}{12\|H(0)\|\|[D,A]\|}\;.
$$
By \eqref{eq:rho2}
$$
\rho\;>\;\frac{8g}{3\kappa}\;>\;\frac{2g(0)}{\kappa}\;.
$$
As $L_{\kappa, \rho}(0)$ is unitary equivalent to 
$$
\widehat{L}_{\kappa, \rho}(0)
\;=\;\begin{pmatrix} \kappa D & A \\  A^* & -\kappa D\end{pmatrix}_\rho\oplus \begin{pmatrix} -\kappa D & A^* \\ A & \kappa D\end{pmatrix}_\rho
\;,
$$
Theorem \ref{theo-SpecLoc} implies that $L_{\kappa, \rho}(0)$ is invertible and 
$$
L_{\kappa, \rho}(0)^2\;\geq\;\frac{g(0)^2}{4}\;\geq\;\frac{(\tfrac{2g}{3})^2}{4}
\;=\;
\frac{g^2}{9}\;.
$$
Therefore
$$
\|L_{\kappa, \rho}(0)^{-1}\|^{-1}\;\geq\;\frac{g}{3}\;.
$$
As $\|L_{\kappa, \rho}(0)-L_{\kappa, \rho}(t)\|<\frac{g}{3}$ by the bound on $\eta$, this implies that $L_{\kappa, \rho}(t)$ is invertible for all $t\in [0,1]$. Therefore
$$
\frac{1}{2} \;\Sig  \left( L_{\kappa, \rho} \right)
\;=\;
\frac{1}{2} \;\Sig  \left( L_{\kappa, \rho}(0) \right)
\;=\;
2\;\Ind (EA|A|^{-1}E+\one-E) \;,
$$
where the last step follows from Theorem~\ref{theo-SpecLoc}.
\hfill$\Box$

\vspace{.2cm}

\noindent{\bf Proof of Proposition \ref{prop-SpecLocApproxConsLaw}.} Here one uses the path
$$
t\in[0,1]\mapsto
L_{\kappa, \rho}(t)
\;=\;
\begin{pmatrix}
H(t) & \kappa \, D \, \sigma_1
\\
\kappa\,\sigma_1 \,D & -H(t)
\end{pmatrix}
\quad
\mbox{ with }
\quad
H(t)
\;=\;
\begin{pmatrix} H_+ & tA \\ tA^* & H_-\end{pmatrix}
\;,
$$
and shows as above that it lies in the invertibles for admissible $(\kappa,\rho)$. Computing the constant signature at $t=1$ and $t=0$, one concludes again similar as in the proof of Proposition~\ref{prop-SpecLocApproxChiral}.
\hfill$\Box$


\end{document}